\documentclass[5p,final]{elsarticle}
\usepackage{amsfonts}
\usepackage{amssymb}
\usepackage{amsmath}  
\usepackage{amsthm}
\usepackage{enumitem} 
\usepackage{geometry}
\usepackage{graphicx}% Include figure files
\usepackage{algorithm}
\usepackage{algcompatible}
\usepackage[colorlinks=true, allcolors=blue]{hyperref}
\usepackage{hyperref}
\usepackage[utf8]{inputenc} 
\usepackage[english]{babel} 
\usepackage{url}
\usepackage{wrapfig}
\usepackage{graphicx}
\usepackage{subfiles}
\usepackage{xfrac}
\usepackage{multirow}
\usepackage{cleveref}
\usepackage{balance}
\usepackage{booktabs}
\usepackage{tikz}
\usetikzlibrary{shapes.geometric, arrows}
\usepackage{tikz-3dplot}
\usepackage[squaren]{SIunits}
\usepackage{caption}
\usepackage[normalem]{ulem}
\usepackage{subfig}
\usepackage{lipsum}
\usepackage{todonotes}
\usepackage{float}
\newtheorem{theorem}{Theorem}

\newcommand{\bc}{\begin{center}}
\newcommand{\ec}{\end{center}}
\newcommand{\be}{\begin{equation}}
\newcommand{\ee}{\end{equation}}
\newcommand{\bea}{\begin{eqnarray}}
\newcommand{\eea}{\end{eqnarray}}
\newcommand{\beq}{\begin{eqnarray*}}
\newcommand{\eeq}{\end{eqnarray*}}
\newcommand{\bv}{\left( \begin{array}{c} }
\newcommand{\ev}{\end{array} \right) }

\newcommand{\R}{\mathbb{R}}
\newcommand{\C}{\mathbb{C}}

\newcommand{\E}{\mathbb{E}}

\newcommand{\unitv}{\hat{\vec e}}
\newcommand{\hata}{\hat {\ve \alpha}}

\newcommand{\argmax}{\mathrm{argmax}}

\newcommand{\diag}{\mathrm{diag}}
\newcommand{\trace}{\mathrm{tr}}

\renewcommand{\vec}[1]{\mathbf{#1}}
\newcommand{\ve}[1]{\boldsymbol{#1}} % for greek letters

\usepackage{natbib}
\begin{document}   
\selectlanguage{english}
\title{Geometric insights into robust portfolio construction}
	\author[1]{Lara Dalmeyer}
	\ead{DLMLAR001@myuct.ac.za}
    \author[1]{Tim Gebbie}%\orcid{0000-0002-4061-2621}
    	\ead{tim.gebbie@uct.ac.za}
	\address[1]{Department of Statistical Sciences, University of Cape Town, Rondebosch 7701, South Africa}
	\date{\today}
	\begin{abstract}
We investigate and extend the result that an $\ve \alpha$-weight angle from unconstrained quadratic portfolio optimisations has an upper bound dependent on the condition number of the covariance matrix. This is known to imply that better conditioned covariance matrices produce weights from unconstrained mean-variance optimisations that are better aligned with each assets expected return. Here we relate the inequality between the $\ve \alpha$-weight angle and the condition number to extend the result to include portfolio optimisations with gearing constraints to provide an extended family of robust optimisations. We use this to argue that in general the equally weighted portfolio is not preferable to the mean-variance portfolio even with poor forecast ability and a badly conditioned covariance matrix. We confirm the distribution free theoretical arguments with a simple Gaussian simulation. 
	\end{abstract}
	\begin{keyword}mean-variance optimization \sep Kantorovich inequality \sep gearing constraints \\
	MSC: 91G10 90C20 62P05
	\end{keyword}
	\maketitle
%	\tableofcontents

% \tableofcontents

% Begin the sections
% ---------------------- %
\section{Introduction}
% ---------------------- %
Using a geometric construction that relates the angle between the vector of expected returns $\ve \alpha$ and portfolio weights $\ve \omega$ determined under an unconstrained mean-variance optimisation,  \citet{goltsandjones2009} were able to quantify the distortions induced by the optimisation process without making direct distributional assumptions about the underlying data generating process. They were able to link this angle directly to the condition number of the covariance matrix by showing that this angle has an upper bound dependent on the condition number. The interpretation is novel because of its geometric construction and the weak distributional assumptions this allows.  When a covariance matrix is ill-conditioned, or close to degenerate, the $\ve \alpha$-weight angle should be very large; whereas well-conditioned matrices ensure the angle remains within tight bounds. This is much more than merely a representation change, because it provides a geometric explanation, why an equally weighted portfolio is not always preferable to a mean-variance solution even in the presence of estimation uncertainty. 

In general portfolio construction, a key question remains to what extent one should drag the optimal bets towards the direction of expected asset returns. In the unconstrained case this is clear because in the mean-variance setting the resulting portfolios will be multipliers, or geared realisations, of the optimal risky portfolio, that are the result of solving a Sharpe ratio maximisation. In the constrained case this appears less clear because of how constraints can dominate the solution space. 

Towards this end we investigate the mathematical relationship between the $\ve \alpha$-weight angle, and the condition of the covariance matrix in more generality. \citet{goltsandjones2009} argued the case for dragging optimal weights towards the direction of $\ve \alpha$ for a family of unconstrained quadratic optimisations. We extend this family of portfolio choice problems to include a gearing constraint and to consider whether their geometric insights are more generally applicable for practical investment management decision making. The gearing constraint defines exactly how much of the portfolio should be invested (whether it be fully invested, partially invested or geared) and is a fundamental portfolio construction constraint in much of the unit-trust, exchange traded fund and the related collective investment scheme industry. This constraint is to be differentiated from a leverage constraint, which relates the size of portfolio bets relative to a level of borrowing. 

The important point in the geometric setting is that the gearing constrained optimal portfolio has a weakened upper bound. \citet{goltsandjones2009} 
relied on the Kantorovich inequality \citep{kant21964}, we argue that this can be weakened because of the impact of the constraint using \cite{bauerhouseholder1960}. Noting that with improved conditioning of the covariance matrix, we can reduce the upper bound on the angle between the optimal risky portfolio and $\ve \alpha$, which would in turn reduce the overall angle between the optimal portfolio and the direction of $\ve \alpha$. To our knowledge, this insight is novel; albeit subtle. 

 Concretely, we show that a minimisation over the total $\alpha$-angle $\varphi$ is equivalent to minimising the angle between the optimal risky portfolio and the direction of expected returns. This then extends the unconstrained result and motivates why the combination of an $\ve \alpha$-angle minimisation is equally useful both in the unconstrained and constrained cases. The key point is then that to minimise the overall angle between the optimal portfolio and the expected returns we can minimise the angle between the optimal risky portfolio and the expected returns. 

Reducing the angle between the weights and expected return vector is not equivalent to the implementation of a return maximisation, or return maximisation constraint. The case of pure return maximisation would result in a portfolio that weighs toward the asset with the greatest expected return, and not the direction of the expected return vector in general. In practice the $\ve \alpha$-weight angle should never be zero as there is always a residual correlation between the assets. In mean variance optimisations there is always a conundrum due to correlation specification error – and then determining the degree of shrinkage that should be adopted to best reflect this. The test of how much one should intervene can only be determined by out-of-period portfolio performance simulation, or with direct distributional assumptions (as demonstrated with simple simulations in Section \ref{sec:simports}). 

However, we provide a geometric explanation of why, when faced with estimation uncertainty using a quadratic optimisation, the direction of returns is still of value even if the quantum of expected returns is not, without making strong {\it a-priori} distributional assumptions. 

At a superficial level this work seems to merely promote the important benefits related to improving the conditioning of the covariance matrix in the out-of-sample performance of an asset portfolio if the estimated covariance matrix is used for decision making -- a widely known idea that is common cause. However, here we specifically promote these benefits in the sense of reducing the angle between the direction of the optimal portfolio in the control space relative to that of the expected asset returns. This is a refined perspective with a sound theoretical motivation. This specificity better explains {\it where} and {\it how} improved conditioning may be beneficial.

Theoretically, the key insight is that it is not merely covariance matrix inversion, hence error maximisation more generally, that is problematic, but rather that the inversion of a poorly conditioned covariance matrix pushes the direction of an optimal portfolio erroneously away from the direction of expected returns. This implies that when improving the conditioning of a covariance matrix this should be done in a way that pulls the optimal portfolio back towards the direction of expected returns -- irrespective of whether or not one may be in the presence of constraints. Merely improving the conditioning may prove to be sub-optimal.

Practically, these insights support the perspective promoted by \citet{ALS2019}, that an equally weighted portfolio is not always preferable to a mean-variance solution. We provide a distribution free argument that even with poor forecast ability and the impact of estimation uncertainty, that there can be an advantage to using mean-variance portfolio theory, over an equally weighted (or any risk only) portfolio construction.

\subsection{Estimation uncertainty}

Quadratic or mean–variance analysis was introduced by Harry Markowitz in his seminal 1952 paper \citep{markowitz1952} to establish what is today known as Modern Portfolio Theory. Within this framework a rational investor selects a portfolio based on the trade-off between estimated expected return and risk. Despite the elegance of this, the practical implementation has a variety of short-comings, with optimised portfolios perceived to to ``behave badly" out of sample. \citet{michaud1989} argues that mean variance optimisation often leads to non-intuitive results with unstable weights, further characterising it as being an``error maximiser".  

Additionally, even if means and variances could accurately represent asset class distributions, the theory assumes both future expected returns, and the covariance matrix of asset returns are known, or even that they can be known.  The presence of estimation error in input parameters unfortunately makes the use of Modern Portfolio Theory in its classical Markowitz formulation impractical -- this is well understood and common cause \citep{Paskaramoorthyetal2021}. The short-comings of mean-variance optimisation has been extensively discussed in the literature, with highlights including the work by \citet{bestandgrauer1991}, \citet{fsp1971}, \citet{jak1980,jak1981}, \citet{zan}, \citet{broadie1993}, \citet{michaud1989}, and  \citet{chopraziemba1993}.

A broad variety of different forms and approaches to robust optimisation are documented in an extensive literature. As a small subset, \citet{cap2002}, \citet{hat2003}, \citet{gai2003}, and \citet{cas2003} have proposed various, and unique, specifications of the uncertainty set of the input parameters. Performance and risk metrics of robust portfolios have been shown to be distinct from \cite{markowitz1952} optimal portfolios. \citet{tak2004} showed that robust portfolios tend to investment in fewer assets and display less turnover over time. Increase in the size of the uncertainty sets, when using robust portfolio optimisation procedures, means they tend to under-perform in the more likely scenarios, but display more reasonable tail behaviour.

In order to ameliorate the effect of estimation errors in the estimates of expected returns, attempts have been made to create better and more stable mean-variance optimal portfolios by utilising expected return estimators that have a better behaviour when used in the context of the mean-variance framework \citep{Paskaramoorthyetal2021}. One of the more common techniques is the utilisation of James-Stein estimators (see \citet{jak1981}) to shrink the expected returns towards the average expected return based on the volatility of the asset and the distance of its expected return from the average. \citet{jorion1985} developed a similar technique that shrinks the expected return estimate towards the minimum variance portfolio. The area of robust statistics (see \cite{cavadini}) has recently been employed to create stable expected return estimates as well. \citet{chopraziemba1993} show that errors in the sample mean estimate have a larger impact on the out-of-sample performance than errors in the sample covariance estimate.

In a similar way to estimating robust expected return inputs, one can also build a more robust process through focusing on the estimation of the covariance matrix, which is inverted in the mean variance optimisation process. Papers investigating different covariance matrix estimation techniques include \citet{DisatnikandBenninga}, \citet{clarkedesilvathorley}, \citet{jagandma} and \citet{bengandholst}.This paper makes use of the methodology employed by \citet{ledoitwolf2004}, who argue that a shrunk covariance matrix should be used instead of the sample covariance matrix as it contains significant estimation error.The shrunk matrix represents a weighted average of the sample covariance matrix, and another structured matrix.

\citet{black1992} combine investor’s views with expected returns to generate more robust portfolios. They use reverse optimisation, using portfolio weights and the covariance matrix as inputs and providing the expected returns as output, avoiding the problem of estimation error in expected returns. 

The Bayesian approach has been widely recommended for dealing with the estimation errors in the sample estimates, which involves eliminating the dependence of the optimisation process on the true parameters by replacing them with a prior distribution of the fund manager’s view. \citet{brown1976} and \citet{kaz2007} both investigate the Bayesian approach.

Other attempts to make improvements on the Markowitz mean-variance optimisation issues also includes the portfolio re-sampling methodology of \citet{michaudt} and \citet{laixingchen2011}. Michaud introduces a statistical re-sampling technique that indirectly considers estimation error by averaging the individual optimal portfolios that result from optimising with respect to many randomly generated expected-return and risk estimates. However, portfolio re-sampling is an {\it ad-hoc} methodology that has pitfalls (see \citet{scherer2002}). \citet{laixingchen2011} propose a solution to a stochastic optimisation problem by extending Markowitz’s mean-variance portfolio optimisation theory to the case where the means and covariances of the asset returns for the next investment period are unknown. 

There are a variety of approaches to shrinking the covariance matrix gap between the largest and smallest eigenvalues, these include optimisation based approaches \citep{PUN2019754} whose performance is measured using Sharpe ratio's and certainty equivalents. Such approaches are novel, but do not directly address the issue of back-test over-fitting \cite{bailey2015} nor do they provide distribution free setting with theoretical bounds for both constrained and unconstrained portfolio's.  

\citet{goltsandjones2009} provided a novel geometric perspective on robust portfolio construction using insights that arise when one separates the direction and magnitude of the portfolio position vector. They make no assumptions about the distribution of the returns and suggest making the covariance matrix better-conditioned by constraining the angle between the vector of optimised weights and expected returns -- the ``alpha" angle. The motivation for implementing a minimisation of the $\ve \alpha$-angle is based on showing that this angle is bound by the condition number of the estimated covariance matrix.

Here we demonstrate this relationship geometrically and mathematically using an extension of the Kantorovich inequality \citep{kant21964,kant11985} provided by \citet{bauerhouseholder1960}; this elegant machinery will be shown as useful in determining why this relationship exists (outlines to both theorems are provided in \ref{app:Thrm1} and \ref{app:Thrm2}).  

The theoretical advantage of the geometric approach is not only the distribution free setting, but the theorems directly motivate the choice of shrinkage in terms of the angle between the portfolio controls and the direction of asset returns, and thus provide a direct and unique approach. This is particularly useful if one is able to predict returns and correlations intermittently in real financial markets.

Inspiration for this idea can be seen in \citet{axelsson1996} where much of the geometric interpretation around matrices is described.

\subsection{The efficient frontier and $\alpha$-weight angle}

Consider a portfolio with $n$ assets, expected return vector $\ve \alpha$, covariance matrix $\Sigma$, and a fixed portfolio mean $\alpha_p$. To find this portfolio on the efficient frontier with asset weights $\ve \theta$, we perform a portfolio variance minimization \citep{markowitz1952,merton1972}:
\begin{equation} \label{eq:0}
 \arg \min_{\ve \theta} \left\{ {\frac{1}{2}\ve \theta'\Sigma \ve \theta} \right\} ~~\mbox{s.t.}~~  \ve \alpha' \ve \theta = \alpha_p.
\end{equation} 
Here we can use the Lagrange method and solve the Kuhn-Tucker conditions to find the optimal portfolio allocation as follows:
\begin{equation} \label{eq:1}
\ve \theta^* = \frac{\alpha_p}{\ve \alpha' \Sigma^{-1} \ve \alpha}\Sigma^{-1} \ve \alpha.
\end{equation}
This is a classical Markowitz-type optimisation. It can be shown that all optimisation solutions where there are no gearing, or long-only constraints, are of similar form. The solutions $\ve\theta \propto \Sigma^{-1}\ve\alpha$ all show that the results are calculated using the inverse of the covariance matrix contracted in the direction of the expected returns.

Various criticisms with regards to  Markowitz portfolios noted that mean-variance optimised portfolios can have weights ($\ve \theta$) that are quite different from the original ``alphas'' ($\ve \alpha$). To quantify this difference one can look at the angle ($\phi$) between the vector of expected returns, $\ve \alpha$, and the vector of optimised weights, $\ve \theta^*$ where $\phi \in [0,\pi/2]$. 

The mathematical formula for the angle is as follows:
\begin{equation} \label{eq:2}
\cos(\phi)=\frac{\ve \alpha^\prime\ve\theta}{\mid{\ve \alpha}\mid\mid{\ve \theta}\mid}=\frac{\ve \alpha^\prime\ve\theta}{\sqrt{\ve \alpha'\ve \alpha}\sqrt{\ve \theta'\ve \theta}}.
\end{equation}
The smaller this angle, the more aligned optimal portfolio weights are to expected $\ve \alpha$'s which would indicate a more reasonable and intuitive optimisation result. A poor optimisation result would be indicated by an almost perpendicular angle close to 90$^\circ$ .

\citet{goltsandjones2009} describe the mathematical relationship between this angle and the condition of the covariance matrix; showing how a well conditioned covariance matrix can ensure a smaller $\ve \alpha$-weight angle (because $\cos(\phi)$ is closer to one), and hence a more reasonable optimisation result. They show how this angle is bounded above zero by the condition number of the covariance matrix:
\begin{equation} \label{eq:goltjones3}
\cos(\phi) \geq 2\sqrt{\frac{\kappa}{(\kappa+1)^2}} ~\mbox{where} ~\kappa =  \frac{\rho_{\mathrm{max}}}{\rho_{\mathrm{min}}}.
\end{equation}
Here the condition number $\kappa$ is the ratio of the largest eigenvalue, $\rho_{\mathrm{max}}$, to the smallest eigenvalue, $\rho_{\mathrm{min}}$. It is important to note that although $\cos(\phi)$ has a lower bound, the actual angle $\phi$ has an upper bound, as $\cos(\phi)$ decreases where $\phi \in [0,\pi/2]$ increases.

Despite the intuitive nature of the result, the mathematical reasoning behind how this relationship came to be in the \citet{goltsandjones2009} paper was not fully described and was only applied to portfolio optimisations that were unconstrained. To clarify, this paper refers to unconstrained optimisations as those described in Table \ref{tab:unconstrainedMV}, despite the fact that optimisation I and II each have a risk and return constraint respectively, and as the paper will show, some constraints are implicit in the problem. The optimisations described in Table \ref{tab:linearcons}, where gearing constraints are explicitly put in place, are referred to as the constrained optimisations.

\begin{table}[h] %\label{tab:MV-cons}
%\scriptsize
\centering
\begin{tabular}{p{0.3cm}ll}
\toprule
\multicolumn{3}{c}{No Gearing Constraint Portfolio Choice} \\
\midrule
\# & Optimisation & Constraints \\
\midrule
\multirow{2}{*}{I} & $\arg \displaystyle\max_{\ve \theta} \left\{ {\ve \theta' \ve \alpha} \right\}$ & $\sigma^2_p=\ve \theta' \Sigma \ve \theta \le \sigma^2_0$ \\ 
& \multicolumn{2}{l}{$\mbox{soln:}~\ve \theta_1^* =\left({ \frac{\sigma_0}{\sqrt{C}}}\right) \Sigma^{-1} \ve \alpha= \sigma_0 \frac{B}{\sqrt{C}}\ve \theta_{\alpha}$} \\
\midrule
\multirow{2}{*}{II} & $\arg \displaystyle\min_{\ve \theta} \left\{ { \frac{1}{2} \ve \theta' \Sigma \ve \theta} \right\}$ & $\alpha_p = \ve \theta' \ve \alpha \ge \alpha_0$ \\ 
& \multicolumn{2}{l}{$\mbox{soln:}~\ve \theta_2^* = \left({\frac{\alpha_0}{C}}\right) \Sigma^{-1} \ve \alpha = \alpha_0 \frac{B}{C} \ve \theta_{\alpha}$}\\
\midrule
\multirow{2}{*}{III} &   $\arg \displaystyle\max_{\ve \theta} \left\{ {\ve \theta' \ve \alpha - \frac{\gamma}{2} \ve \theta' \Sigma \ve \theta} \right\}$ & \\ 
&  \multicolumn{2}{l}{$\mbox{soln:}~\ve \theta_3^* = \frac{1}{\gamma} \Sigma^{-1} \ve \alpha = \frac{B}{\gamma} \ve \theta_{\alpha}$}\\
\midrule
\multirow{2}{*}{IV} & $\arg \displaystyle\max_{\ve \theta} \left\{ \frac{\ve \theta' \ve \alpha}{\sqrt{\ve \theta' \Sigma \ve \theta}}  \right\} $ & \\
&  \multicolumn{2}{l}{$\mbox{soln:}~\ve \theta_4^* = \left({\frac{\sigma_p}{\alpha_p}}\right) \Sigma^{-1} \ve \alpha =  \left({\frac{g_0}{B}}\right) \Sigma^{-1} \ve \alpha = g_0  \ve \theta_{\alpha}$}\\
\bottomrule
\end{tabular}
\caption{Some classical unconstrained portfolio optimisation problems from \cite{goltsandjones2009}. Here portfolios employ no explicit gearing constraint, and positions may be unconstrained long/short. Here I is a return maximisation problem with risk bound, II is a risk minimisation problem with a return bound, III is a conventional mean-variance optimisation, and IV is the Sharpe ratio optimisation. In III $\gamma$ is the inverse risk aversion parameter. In IV the risk free rate $r_f$ is assumed to be zero. Here $A = {\vec 1' \Sigma^{-1}\vec 1}$,$B = {\ve \alpha' \Sigma^{-1}\vec 1} = \vec 1\Sigma^{-1}\ve \alpha$, $C = {\ve \alpha'\Sigma^{-1}\ve \alpha}$ and $D = AC - B^2$. All the portfolio's are implicitly geared versions of the optimal risky portfolio $\ve \theta_{\alpha}$. Optimisation IV requires an explicit choice of $g_0$ to set the magnitude of the optimal portfolio which is typical set to one: $g_0=1$. The optimal portfolio controls $\ve \theta^*$ are in the last column for each optimizations in the table. The subscript $i$ on the solution $\ve \theta_i^*$ is the number of the optimisation in question.}\label{tab:unconstrainedMV}
\end{table}

\subsection{Unconstrained $\alpha$-weight angle} \label{ssec:unconstrained}

\begin{theorem}{Kantorovich Inequality \citep{kant21964}} \label{thm:kant}\\
Let $T\in M_n(R)$ be a symmetric positive definite matrix with eigenvalues $0\leq \rho_n \leq \ldots \leq \rho_1$. Then for all $\vec v\in \R^n\backslash\{0\}$,
\begin{equation} \label{eq:kant1}
\frac{\left({\vec v'\vec v}\right)^2}{\left({\vec v'T\vec v}\right)\left({\vec v'T^{-1}\vec v}\right)} \geq  \frac{4 \rho_1\rho_n}{\left(\rho_1+\rho_n\right)^2}.
\end{equation}
\end{theorem} 

We first motivate the original unconstrained result in our Eqn.(\ref{eq:goltjones3}) \citep{goltsandjones2009} using Theorem \ref{thm:kant} the Kantorovich inequality \citep{kant21964,Householder1965,kant11985} -- this is a special case of the Cauchy-Schwarz inequality. We provide an outline of the proof of Theorem \ref{thm:kant} following \cite{Householder1965} in \ref{app:Thrm1}. Here we will use equations \ref{eq:kant1} (From Theorem \ref{thm:kant}) and \ref{eq:1}, and that any positive semi-definite matrix $\Sigma$ can be decomposed into the product of two symmetric matrices $S$. 

We start with optimal unconstrained portfolio solution from Eqn. \ref{eq:1}:
\begin{equation}\label{eq:riskminsoln}
    \ve \theta^* \propto \Sigma^{-1}\ve\alpha,
\end{equation}
This is substituted into the $\alpha$-weight angle using Eqn. (\ref{eq:2})
\begin{equation} \label{eq:5}
\cos(\phi)=\frac{\ve\alpha'(\Sigma^{-1}\ve\alpha)}{\sqrt{\ve\alpha'\ve\alpha}\sqrt{(\Sigma^{-1}\ve\alpha)'(\Sigma^{-1}\ve\alpha)}},
\end{equation}
where, because a symmetric matrix is equivalent to its own transpose:
\begin{equation} \label{eq:7}
\cos(\phi)=\frac{\ve\alpha'\Sigma^{-1}\ve\alpha}{\sqrt{\ve\alpha'\ve\alpha}\sqrt{\ve\alpha'\Sigma^{-2}\ve\alpha}}. 
\end{equation}
Using that an inverse covariance matrix can be factored in terms of a symmetric matrix $S$, where $\Sigma^{-1} = S' S$, Eqn.(\ref{eq:7}) becomes:
\begin{equation} \label{eq:8}
\cos(\phi)= \frac{\ve\alpha'S'S\ve\alpha}{\sqrt{\ve\alpha'\ve\alpha}\sqrt{\ve\alpha'S'SS'S\ve\alpha}}.
\end{equation}
Now let $\ve x=S\ve \alpha$ so that $\ve \alpha = S^{-1} \ve x$,  $\ve x'=\ve \alpha'S$ and  $\ve \alpha'=\ve x'S^{-1}$, and substitute these into Eqn.(\ref{eq:8}) to find:
\begin{equation} \label{eq:9}
\cos(\phi)= \frac{\ve x' \ve x}{\sqrt{\ve x'\Sigma \ve x}\sqrt{\ve x'\Sigma^{-1}\ve x}}.
\end{equation}
The right hand side of Eqn.(\ref{eq:9}) is the square root of the left hand side of the Kantorovich inequality from Eqn. (\ref{eq:kant1}) thereby implying that this is bounded: 
\begin{equation} \label{eq:10}
\cos^2(\phi) = \frac{(\ve x'\ve x)^2}{(\ve x'\Sigma \ve x)(\ve x'\Sigma^{-1}\ve x)} \geq  \frac{4 \rho_1\rho_n}{\left(\rho_1+\rho_n\right)^2},
\end{equation}
where $\Sigma$ has ordered eigenvalues $0\leq \rho_n \leq \ldots \leq \rho_1$.

Following \citet{goltsandjones2009} with the condition number of the covariance matrix: $\kappa = \sfrac{\rho_1}{\rho_n}$, we can then show that an upper bound to the $\ve \alpha$-weight angle exists, and we recover Eqn.(\ref{eq:goltjones3}):
\begin{equation} \label{eq:12}
\cos^2(\phi) \geq \frac{4 \rho_1\rho_n}{(\rho_1+\rho_n)^2} =  \frac{4\kappa}{{(\kappa+1)}^{2}}~ \mathrm{where}~ \kappa = \frac{\rho_{\mathrm{max}}}{\rho_{\mathrm{\min}}}
\end{equation}
This is the result given in \cite{goltsandjones2009}, but explicitly derived here to motivate the form of our constrained extension. This shows that an improved condition number of the covariance matrix can improve the robustness of the optimisation procedure, but only when constraints, such as that with gearing, a full investment constraint, or a long-only portfolio constraint, are ignored. 

The idea is that as the conditioning of the covariance deteriorates the $\alpha$-angle necessarily increases. With improved conditioning, when there are no constraints, then the optimal portfolio weights can more easily align with the directions of the asset expected returns (See Figure \ref{fig:geometry1}) and so improve the out-of-sample portfolio performance. This is because as the optimisation aligns with the direction of the expected returns, and if the expected returns adequately reflect future scenario's, then this implies improved out-of-sample portfolio performance.

\begin{figure}[ht!]
   \centering
        %\resizebox{\columnwidth}{!}{
        \tdplotsetmaincoords{70}{110}
        \begin{tikzpicture}[scale=3,tdplot_main_coords]
            
            \coordinate (O) at (0,0,0);
            %coordinate system
            \draw[thick,->] (-1,0,0) -- (1,0,0) node[anchor=north east]{$\omega_1$};
            \draw[thick,->] (0,-0.5,0) -- (0,1,0) node[anchor=north west]{$\omega_2$};
            \draw[thick,->] (0,0,-0.5) -- (0,0,1) node[anchor=south]{$\omega_3$};
            
            % alpha vector
            \tdplotsetcoord{A}{1.1}{65}{50};
            %draw a vector from origin to point (P)
            \draw[very thick, -stealth,color=red] (O) -- (A) node[anchor=south]{$\ve \alpha$};
            %draw projection on xy plane, and a connecting line
            \draw[dashed, color=red] (O) -- (Axy);
            \draw[dashed, color=red] (A) -- (Axy);
            
             % optimal portfolio
            \tdplotsetcoord{T}{0.9}{60}{120};
            %draw a vector from origin to point (P)
            \draw[very thick, -stealth,color=black] (O) -- (T) node[anchor=south]{$\ve \theta^*$};
            %draw projection on xy plane, and a connecting line
            \draw[dashed, color=black] (O) -- (Txy);
            \draw[dashed, color=black] (T) -- (Txy);
            
            % optimal risky portfolio
            \tdplotsetcoord{G}{1.3}{60}{120};
            %draw a vector from origin to point (P)
            \draw[-stealth,color=red] (O) -- (G) node[anchor=south]{$\ve \theta_{\alpha}$};
            %draw projection on xy plane, and a connecting line
            \draw[dashed, color=red] (O) -- (Gxy);
            \draw[dashed, color=red] (G) -- (Gxy);
            
            % draw the arc
            \tdplotdefinepoints(0,0,0)(-1,2,1.2)(2,3,1.4);
            \tdplotdrawpolytopearc[thick, <->,  -stealth]{1}{anchor=west}{$\phi$};
        \end{tikzpicture}
    %}
    \caption{Unconstrained geometry of the $\ve \alpha$-angle where $\ve \alpha$ are the asset expected returns, $\ve \theta^*$ some optimal portfolio weights, $\ve \theta_{\alpha}$ is the optimal risky portfolio, and $\phi$ is the $\ve \alpha$-angle. In the unconstrained case $\ve \theta_{\alpha}$ and $\ve \theta^*$ are aligned because the optimal portfolio is just a scaled version of the optimal risky portfolio. The optimal portfolio is in the direction of the vector $\Sigma^{-1} {\ve \alpha}$ and hence does not in general align with the $\ve \alpha$ direction: $\hat {\ve \alpha}$. If the covariance matrix $\Sigma$ is shrunk to the identity matrix then all three vectors will align.}
    \label{fig:geometry1}
\end{figure}

\section{The constrained $\alpha$-weight angle} \label{sec:constrained}

Many funds have gearing restrictions and may not be outright short an instrument {\it i.e.} $\ve \theta \ge \vec 0$. A portfolio optimisation enforcing a long only constraint often has the problem that it cannot ensure that closed form solutions can be found. It is similarly problematic to include margin and diversity constraints. A margin constraint is required when investing in futures, or where some sort of deposit proportional to the investment made is required \citep{GALLUCCIO1998449,GABOR1999222}: {\it i.e.} $\sum_i m_i|\ve \theta_i| = 1$, for $m_i$ the margin account required for $i$-th asset. When the level of margining is the same for all the assets this can be written as a leverage constraint  of the form: $\vec 1' |\ve \theta|=\ell_0$.

However, it is possible to get closed form solutions for a gearing constrained optimisation; using the constraint $\ve \theta' \vec 1 = g_0$ for some gearing parameter $g_0$. A fully invested portfolio with constraint $\vec 1' \ve \theta =1$ is then a special case. These can be implemented as: a return maximisation with a linear and quadratic constraint, a portfolio variance minimisation with two linear constraints, a mean-variance optimisation with a linear constraint, or a Sharpe ratio maximisation with a linear gearing constraint (See Table \ref{tab:linearcons}).  

\subsection{Upper bound on the $\ve \alpha$-weight angle} \label{sec:existance}

\begin{theorem}{Bauer and Householder Inequality \citep{bauerhouseholder1960}} \label{thm:bah}\\
Let $\vec p, \vec q \in \R^{n}$ such that
$\frac{\vec p'\vec q}{\|\vec p\|_2\|\vec q\|_2}\geq\cos(\psi)$ with $0\geq\psi\ge\frac{\pi}{2}$. Then
\begin{equation} \label{eq:39}
\frac{\left(\vec p'\vec q\right)^2}{\left(\vec p'T \vec p \right)\left(\vec q'T^{-1}\vec q\right)} \geq \frac{4}{\kappa_{\psi}+2+\kappa_{\psi}^{-1}}
\end{equation}
where $T$ is the same positive semi-definite matrix used in Theorem \ref{thm:kant} and
\begin{equation} \label{eq:bah-kappa}
    \kappa_{\psi}=\frac{\rho_1}{\rho_n}\frac{1+\sin(\psi)}{1-\sin(\psi)}
\end{equation}
is a function of the angle between the vectors $\vec p$, and $\vec q$.
\end{theorem}

To prove that the upper bound on the $\ve \alpha$-weight angle exists for linearly constrained mean-variance optimisations, we make use of the result derived by \citet{bauerhouseholder1960} who extended the Kantorovich inequality (Theorem \ref{thm:kant}) to Theorem \ref{thm:bah}. A direct by construction proof for this is investigated by \citet{huangandzhou}. For completeness we rather outline the proof of Theorem \ref{thm:bah} following the original approach of \cite{bauerhouseholder1960} in \ref{app:Thrm2}. We note that in Theorem \ref{thm:bah}, and in particular from Eqn. \ref{eq:bah-kappa}, that $\kappa_{\psi}$ is unbound above and bound by $\kappa$ below: $\kappa_{\psi} \in [\kappa,\infty]$ and hence we have that $\kappa_{\psi}\ge \kappa$. This condition is weaker than that of Theorem \ref{thm:kant}:
\begin{equation}
     \frac{4 \kappa}{(\kappa + 1)^2}\ge \frac{4\kappa_{\psi}}{(\kappa_{\psi} + 1)^2} ~\mathrm{where}~ \kappa = \frac{\rho_{\mathrm{max}}}{\rho_{\mathrm{min}}},
\end{equation}
because $\kappa_{\psi} \ge \kappa$ for $0\geq\psi\ge\frac{\pi}{2}$.

\subsection{Constrained $\ve \alpha$-angle}

Following from the definition of the $\alpha$-weight angle from Eqn.
(\ref{eq:2}) we can again find an expression for the angle, however this angle is now that related to the constrained optimisation angle $\varphi$ from $\ve \alpha' \ve \theta^*=|\ve \alpha||\ve \theta^*| \cos(\varphi)$ using that $\ve \theta^* = \Sigma^{-1}\vec z$ represents some general linear solution of the constrained problem where in general $\vec z \neq \ve \alpha$ ({\it e.g.} using Eqn.(\ref{eq:gearingz}), or equivalently use $\vec w$ from Eqn.(\ref{eq:gearingw})), into the $\alpha$-weight angle Eqn.(\ref{eq:2}) for the new angle (say) $\varphi$:
\begin{equation}
\cos(\varphi)=\frac{\ve \alpha' \ve \theta^*}{\sqrt{\ve \alpha\ \ve \alpha}\sqrt{{\ve \theta^*}'\ve \theta^*}}=\frac{\ve \alpha'\Sigma^{-1}\vec z}{\sqrt{\ve \alpha'\ve \alpha}\sqrt{\vec z'\Sigma^{-2}\vec z}}.
\end{equation}
Now, as performed in equations (\ref{eq:7}) to (\ref{eq:9}) we exploit the properties of the positive semi-definite matrix $\Sigma^{-1} = S'S$ to make the substitutions: $\vec x=S \ve \alpha$ and $\vec y=S \vec z$. The desired result then follows from Theorem \ref{thm:bah}:
\begin{equation} \label{eq:37}
\cos(\varphi)= \frac{\vec x'\vec y}{\sqrt{\vec x'\Sigma \vec x}\sqrt{\vec y'\Sigma^{-1}\vec y}} \ge \sqrt{\frac{4 \kappa_{\psi}}{(\kappa_{\psi} + 1)^2}}.
\end{equation}
If, for some angle $\psi$, where $\psi \in[0,\frac{\pi}{2}]$ and $\psi<\varphi$:
\begin{equation} \label{eq:bahangle}
    \cos^2(\varphi) \ge \frac{4 \kappa_{\psi}}{(\kappa_{\psi} + 1)^2} ~\mbox{where}~ \kappa_{\psi}=\frac{\rho_{\mathrm{max}}}{\rho_{\mathrm{min}}}\frac{(1+\sin(\psi))}{(1-\sin(\psi))}. %\nonumber
\end{equation}
Then the $\ve \alpha$-weight angle is remains bound above by the condition number of the covariance matrix: $\kappa$. This extends the geometric interpretation to include linear portfolio constraints, because the $\ve \alpha$-angle is bounded below. Here with more generality, and thus as a weakened condition, to again imply that poor conditioning will limit our ability to align the portfolio weights with the expected returns, but again, without making explicit distributional assumptions about the data generating process.

There is now a tension between the linear constraint, the alignment of the optimal control with the direction of the expected returns, and the impact of the conditioning that would push the optimal portfolio out of alignment. This more complex tension is visualised in Figure (\ref{fig:geometry2}) and we now need to include the role of the global minimum variance  (GMV) portfolio $\ve \theta_0$. The inclusion of the global minimum variance portfolio then introduces a dependency on the equally-weighted portfolio when correlations are negligible. 

%tikz
\begin{figure}[ht]
    \centering
    %\sidecaption
    %\resizebox{\columnwidth}{!}{
        \tdplotsetmaincoords{70}{110}
        \begin{tikzpicture}[scale=3,tdplot_main_coords]
            
            \coordinate (O) at (0,0,0);
            %coordinate system
            \draw[thick,->] (-1,0,0) -- (1,0,0) node[anchor=north east]{$\omega_1$};
            \draw[thick,->] (0,-1,0) -- (0,1,0) node[anchor=north west]{$\omega_2$};
            \draw[thick,->] (0,0,-0.5) -- (0,0,1) node[anchor=south]{$\omega_3$};
            
            % alpha vector
            \tdplotsetcoord{A}{0.9}{50}{240};
            %draw a vector from origin to point (A)
            \draw[very thick, -stealth,color=red] (O) -- (A) node[anchor=south]{$\ve \alpha$};
            %draw projection on xy plane, and a connecting line
            \draw[dashed, color=red] (O) -- (Axy);
            \draw[dashed, color=red] (A) -- (Axy);
            
             % optimal portfolio
            \tdplotsetcoord{T}{1.1}{50}{320};
            %draw a vector from origin to point (T)
            \draw[very thick, -stealth,color=black] (O) -- (T) node[anchor=south]{$\ve \theta^*$};
            %draw projection on xy plane, and a connecting line
            \draw[dashed, color=black] (O) -- (Txy);
            \draw[dashed, color=black] (T) -- (Txy);
            
            % arc 1
            %\tdplotdefinepoints(0,0,0)(Tx,Ty,Tz)(Ax,Tx,Ty);
            %\tdplotdrawpolytopearc[thick]{0.5}{anchor=west}{$\theta$}
            
            % global minimum variance portfolio
            \tdplotsetcoord{G}{1}{65}{35};
            %draw a vector from origin to point (G)
            \draw[-stealth,color=blue] (O) -- (G) node[anchor=west]{$\ve \theta_0$};
            %draw projection on xy plane, and a connecting line
            \draw[dashed, color=blue] (O) -- (Gxy);
            \draw[dashed, color=blue] (G) -- (Gxy);
            
            % optimal risky portfolio
            \tdplotsetcoord{R}{1.3}{50}{220};
            %draw a vector from origin to point (R)
            \draw[-stealth,color=red] (O) -- (R) node[anchor=south]{$\ve \theta_{\alpha}$};
            %draw projection on xy plane, and a connecting line
            \draw[dashed, color=red] (O) -- (Rxy);
            \draw[dashed, color=red] (R) -- (Rxy);
            
             % draw the arc
            \tdplotdefinepoints(0,0,0)(1.4,-2,2)(-1.2,-2,2);
            \tdplotdrawpolytopearc[thick, <->,  -stealth]{0.7}{anchor=north west}{$\varphi$};
        \end{tikzpicture}
    %}
    \caption{Constrained geometry of the $\ve \alpha$-angle where $\ve \alpha$ are the asset expected returns, $\ve \theta^*$ some optimal portfolio weights, $\ve \theta_0$ is the Global Minimum Variance (GMV) portfolio, $\ve \theta_{\alpha}$ the optimal risky portfolio, and $\varphi$ the $\ve \alpha$-angle for some choice of the vector between the optimal portfolio and $\ve \alpha$. The control space gives a vector $\ve \omega \in \R^3$, in general our control space is $\R^n$. Here the gearing constraint $\ve \theta_{\alpha}$ no longer aligns in general with $\ve \alpha$ because the constraints impose a dependency on the global minimum variance portfolio $\ve \theta_0$. }
    \label{fig:geometry2}
\end{figure}

\section{The linear gearing constraint} \label{sec:linear}

We now explicitly consider the functional form of the vector $\vec z$ in the optimal solution $\ve \theta^* = \Sigma^{-1} \vec z$ (and by extension $\vec w$). Consider the optimisations in Table \ref{tab:unconstrainedMV} but now include the gearing restriction:
\begin{equation}\label{eq:gearingcon}
\vec 1' \ve \theta = g_0.
\end{equation}
Here gearing is used broadly in the sense of the ratio of exposure of the investment fund relative to the underlying equity. As explained before, we differentiate this from the concept of leverage, which we use as a measure of the amount we have borrowed to gain exposure. Practically we will consider leverage to be measured by $\ell_0 = \vec 1' |\ve \theta|$. This is important in order to differentiate the type of constraint we will impose in the fund management problems we consider and the limitations of this. The gearing constraint will not in general limit the extent to which bets are long/short within the fund.

Concretely, we consider four unconstrained optimisations, those in Table \ref{tab:unconstrainedMV}, and the four gearing constrained optimisations as shown in Table \ref{tab:linearcons}. We follow the work of \citet{ingersoll1987} and \citet{merton1972} who investigate and derive the results of optimal mean variance portfolios in detail. In our case, we will only need to investigate two portfolio choice problems in detail. First, that of minimising the portfolio variance and targeting the return with a gearing constraint. Second, by finding a mean-variance optimal portfolio with a risk aversion parameter $\gamma$ with the same gearing constraint. Using these solutions we show that their $\ve \alpha$-angles are bounded above (because the cosine is bounded below) by some function of the condition number. We then extend the argument made by \citet{goltsandjones2009} that the dragging optimal bets towards the direction of $\ve \alpha$ is limited by poor conditioning, albeit in a weakened form. These optimisations are common cause, but how they are used here is novel. 

Optimisation V and VIII in Table \ref{tab:linearcons}, respectively the return maximisation with a linear and quadratic constraint, and a Sharpe ratio maximisation with a linear leverage constraint, both trivially have optimal portfolios that are multiplies of the risky portfolio $\ve \theta_{\alpha}$. This means that Theorem \ref{thm:kant} is valid for optimisations I,II,III,VI,V and VIII. This means that we only need to generalise and weaken the condition number bound for optimisation VI and VII (See Figure \ref{fig:efffront}(b)).

\begin{table}[h]%\label{tab:MV-cons} 
%\sidecaption
\centering
\begin{tabular}{p{0.4cm}ll}
\toprule
\multicolumn{3}{c}{Gearing Constrained Portfolio Choice} \\
\midrule
\# & Optimisation & Constraints \\
\midrule
\multirow{2}{*}{V} & $\arg \max_{\ve\theta} \left\{ {\ve \theta' \ve \alpha} \right\}$ & $\begin{matrix}\sigma_p^2 = \ve \theta' \Sigma \ve \theta \le \sigma_0 \\ \vec 1' \ve \theta = g_0\end{matrix}$ \\
 & \multicolumn{2}{l}{$\mbox{soln:}~\ve \theta_5^* = g_0 \ve \theta_{\alpha}$}  \\
 \midrule
\multirow{2}{*}{VI} & $\arg\displaystyle\min_{\ve\theta} \Bigg\{ {\frac{1}{2}\ve \theta'\Sigma\ve \theta} \Bigg\}$ & $\begin{matrix}\alpha_{_P}= \ve \alpha' \ve \theta \ge \alpha_0 \\ \vec 1' \ve \theta = g_0\end{matrix}$ \\
 & \multicolumn{2}{l}{$\mbox{soln:}~\ve \theta_6^* =  (g_0  - \omega) \ve \theta_0  + \omega \ve\theta_{\alpha},~\omega = \left[{\frac{B}{D}( \alpha_0 A -g_0 B)}\right]$.}  \\
 \midrule
\multirow{2}{*}{VII} & $\arg\displaystyle\max_{\ve \theta} \left\{ { \ve \theta'\ve  \alpha - \cfrac{\gamma}{2} \ve \theta'\Sigma \ve \theta }\right\}$ & $\vec 1' \ve \theta = g_0.$ \\
    & \multicolumn{2}{l}{$\mbox{soln:}~\ve \theta_7^* = (g_0 -  \mu) \ve \theta_0 + \mu \ve \theta_{\alpha},~\mu = \left[{\frac{1}{\gamma} B}\right]$} \\
     \midrule
\multirow{2}{*}{VIII} & $\arg\displaystyle\max_{\ve \theta} \Bigg\{ \frac{\ve \theta'\ve  \alpha}{\sqrt{\ve \theta'\Sigma \ve \theta}}\Bigg\}$ & $\vec 1' \ve \theta = g_0.$ \\
    & \multicolumn{2}{l}{$\mbox{soln:}~\ve \theta_8^* = g_0 \ve \theta_{\alpha}$} \\
\bottomrule
\end{tabular}
\caption{Four gearing constrained portfolio optimisation problems that can be extend into the geometric framework of  \cite{goltsandjones2009} (See Table \ref{tab:unconstrainedMV}). Here $A = {\vec 1' \Sigma^{-1}\vec 1}$,$B = {\ve \alpha' \Sigma^{-1}\vec 1} = \vec 1\Sigma^{-1}\ve \alpha$, $C = {\ve \alpha'\Sigma^{-1}\ve \alpha}$ and $D = AC - B^2$. The GMV portfolio is $\ve \theta_0=\Sigma^{-1} \vec 1/A$ and the optimal risky portfolio is $\ve \theta_{\alpha} = \Sigma^{-1} \ve \alpha /B$. The subscript $i$ on the solution $\ve \theta_i^*$ is the number of the optimisation in question.}\label{tab:linearcons}
\end{table}

\subsection{Constrained minimum variance portfolio (VI)} \label{ssec:minriskgearing}

We want to find an investment portfolio with minimum variance and a targeted return with the additional gearing constraint, hence we want to solve for:
\begin{equation}\label{eq:optgearing}
    \arg \min_{\ve\theta} \left\{ {\frac{1}{2}\ve \theta'\Sigma\ve \theta} \right\} \mathrm{ s.t. }~~  \ve \alpha' \ve \theta = \alpha_p \mbox{ and }  \vec 1' \ve \theta = g_0.
\end{equation}
This is optimisation VI in Table \ref{tab:linearcons}. This problem can be reduced to a Lagrangian with two Lagrange multipliers $\ve \lambda = (\lambda_1,\lambda_2)$:
\begin{equation}\label{eq:L}
L =  \frac{1}{2} \ve \theta'\Sigma\ve \theta - \ve \lambda' \left[ {\begin{matrix} \ve \alpha' \ve \theta - \alpha_p \\ \vec 1' \ve \theta - g_0 \end{matrix} } \right].
\end{equation}
Solving for $\ve \theta$ in terms of the Lagrange multipliers $\ve \lambda$ from Eqn.(\ref{eq:L}) by solving the Kuhn-Tucker equations to find:
\begin{equation}\label{eq:optcontrols}
    \ve \theta^*_6 = \Sigma^{-1}\lambda_1\ve\alpha + \Sigma^{-1}\lambda_2\vec1.
\end{equation}
Then with some algebra and using the convenient substitutions: $A = \vec 1' \Sigma^{-1}\vec 1$, $B = \ve \alpha' \Sigma^{-1}\vec 1 = \vec 1\Sigma^{-1}\ve \alpha$, $C = \ve \alpha'\Sigma^{-1}\ve \alpha$ and $D = AC - B^2$. Here A is the variance of the global minimum variance portfolio, B that of the optimal risky portfolio, $\sqrt{C}$ is the Sharpe ratio, and D is positive by the Cauchy-Schwarz inequality because we have assumed that $\Sigma$ is non-singular and all assets do not  have  the  same  means. We can eliminate the two Lagrange multipliers to find the optimal portfolio control:
    \begin{equation}\label{eq:mvgearing}
   \ve \theta^*_6 = \frac{g_0}{D}\Sigma^{-1} (C\vec 1 - B \ve\alpha) + \frac{\alpha_p}{D} \Sigma^{-1} (A \ve\alpha - B \vec 1).
\end{equation}
We can define two new portfolios $\vec a$ and $\vec b$:
\begin{equation}\label{eq:2fund}
    \ve \theta^*_6 = g_0 \vec a + \alpha_p \vec b.
\end{equation}
The optimal portfolio control can then be written in the form of a two fund separation theorem where the portfolios are weighted by a combination of gearing $g_0$ and expected portfolio return $\alpha_p$ weighted portfolios, we might naively anticipate that fixing the level of returns may fix the gearing if the solution is unique -- this is not the case. In the return maximisation (optimisation VI) when we set the portfolio variance we will set the gearing because the constraint is quadratic. That is no longer the case with risk minimisation.

To more intuitively understand the portfolio, we can also split the optimal portfolio into the global minimum variance portfolios $\ve \theta_0 = \Sigma^{-1}\vec 1/A$ and optimal risk portfolios $\ve \theta_{\alpha} = \Sigma^{-1} \ve \alpha/B$. Arranging terms in Eqn.(\ref{eq:mvgearing}):
\begin{equation}\label{eq:mvgearing3}
   \ve \theta^*_6 =  (g_0 C -\alpha_p B)\frac{A}{D} \ve \theta_0  + (\alpha_pA-g_0B)\frac{B}{D}\ve\theta_{\alpha}.
\end{equation}
This is convenient because we know that $\vec 1' (\ve \theta_0 + \ve \theta_{\alpha})=0$, $\vec 1' \ve \theta_0 =1$, and that $\vec 1' \ve \theta_{\alpha} =1$. The combination is a cash neutral portfolio whose weights sum to zero, and the individual portfolios are fully invested. We can factor out the geared global minimum risk portfolio to find 
\begin{equation} \label{eq:minriskgear5}
    \ve \theta^*_6 = g_0 \ve \theta_{0} + \left[{\frac{B}{D} \left( {g_0 B - \alpha_p A} \right)}\right] (\ve \theta_0 - \ve \theta_{\alpha}).
\end{equation}
We see that the resulting portfolio is a combination of a leveraged global minimum risk portfolio and a geared cash neutral portfolio. With some weight $\omega = \frac{B}{D}(g_0 B - \alpha_p A)$ this can be written as: $\ve \theta^* = (g_0  - \omega) \ve \theta_0 + \omega \ve \theta_{\alpha}$. This is the form we will use more generally as in optimisation VI in Table \ref{tab:linearcons}.

If we set the level of returns we notice that this has no effect on the overall gearing of the portfolio as long as it is part of the set of feasible solutions. It will just change our exposure to the cash neutral portfolio. This does not imply that the leverage $|\ve \theta|$ is fixed as this changes with the portfolio return target changes, even for a set level of gearing, because our long/short bets in the cash neutral portfolio will then change. However, there is a minimum return associated with the set of global minimum variance portfolios generated by the different levels of gearing. 

The minimum-variance set can be found using $\sigma_p^2 = {{\ve \theta}^*}' \Sigma \ve \theta^*$ and Eqn.(\ref{eq:optcontrols}):
\begin{equation}\label{eq:portvar}
\sigma_p^2 = \lambda_1 {{\ve \theta_6}^*}' \ve \alpha + \lambda_2 {\ve {\theta_6}^*}' \vec 1 = \lambda_1 \alpha_p + \lambda_2 g_0.
\end{equation}
Or equivalently from Eqn.(\ref{eq:2fund}):
\begin{equation}\label{eq:31}
\sigma^2_p = \frac{1}{D} \left({\alpha_p^2 A -2g_0 \alpha_p B + g_0^2 C}\right).
\end{equation}
We can find the set of global minimum variance portfolios with gearing.

First, find the expected return at the inflection point where the minimum variance portfolio is found:
\begin{equation}
   \frac{\partial \sigma^2_p}{\partial \alpha_p}  = \frac{1}{D} \left({2\alpha_p A - 2g_0B} \right) = 0.
\end{equation}
Then find the target return $\ve \alpha_p$ in terms of the gearing, asset returns and asset covariance: 
\begin{equation} \label{eq:minriskalpha}
    \alpha_p = g_0 \frac{B}{A}.
\end{equation}
We can substitute this into Eqn.(\ref{eq:mvgearing3}) to then verify that this in fact just the geared global minimum variance portfolio:
\begin{equation} \label{eq:minriskgear6}
    \ve \theta^*_6 = g_0 \ve \theta_0,
\end{equation}
Eqn.(\ref{eq:minriskalpha}) can also be used to eliminate the return target from Eqn. (\ref{eq:31}) to find the variance at the point of global minimum variance with gearing: $\sigma_p^2 ={g_0^2}/{A}$.

We would like to now consider how the lower bound in Theorem \ref{thm:kant} is weakened. Rewrite Eqn.(\ref{eq:mvgearing}):
\begin{equation}\label{eq:mvgearing8}
   \ve \theta^*_6 =  \Sigma^{-1} \left[ {(g_0 C -\alpha_p B)\frac{1}{D} \vec 1  + (\alpha_p A-g_0 B)\frac{1}{D}\ve \alpha } \right].
\end{equation}
This can be written in the form of $\ve \theta^* \propto \Sigma^{-1} \vec z$ where $\vec z$ represents any vector of the same dimension as $\ve \alpha$:
\begin{equation} \label{eq:gearingz}
\vec z = {(g_0 C -\alpha_p B)\frac{1}{D} \vec 1  + (\alpha_p A-g_0 B)\frac{1}{D}\ve \alpha }.
\end{equation}
Here $\vec z$ is the same dimension as $\ve \alpha$. By Theorem \ref{thm:bah} we then know the angle between this portfolio and the $\ve \alpha$ portfolio has an upper bound dependent on the condition number of the estimated covariance matrix. It is still advantageous to tilt the gearing constrained optimal portfolio towards $\ve \alpha$ by improving the conditioning of the covariance matrix while maintaining the constraint. 

With the choice of Eqn.(\ref{eq:minriskalpha}) we trivially have that $\vec z = \vec 1$. This would fix the angle between $\ve \alpha$ and the optimal portfolio. This means that the constraint will fix the tilt of the optimal portfolio relative to the $\ve \alpha$ when selecting for the global minimum variance portfolio on the efficient frontier. This means that we cannot naively tilt the optimal portfolio towards $\ve \alpha$ if we have any form of gearing because this would break the constraint. We can however tilt the risky optimal portfolio towards $\ve \alpha$.

\subsection{Constrained mean-variance optimal portfolio (VII)} \label{ssec:meanvariancegearing}

We now consider the relationship between the gearing constrained minimum risk portfolio from Eqn.(\ref{eq:optgearing}) and the gearing constrained mean-variance optimal portfolio:
\begin{align}\label{eq:mv1}
    \argmax_{\ve \theta} \left\{ { \ve \theta'\ve  \alpha - \cfrac{\gamma}{2} \ve \theta'\Sigma \ve \theta }\right\} \quad \mbox{s.t} \quad \vec 1' \ve \theta = g_0.
\end{align}
This is optimisation VII in Table \ref{tab:linearcons}. The optimal control can be found by solving the Kuhn-Tucker equations to find the mean-variance optimal portfolios, Eqn.(\ref{eq:mvgearing}):
\begin{align}\label{eq:mv3}
    \ve \theta^{*}_7 = \left({g_0 - \frac{1}{\gamma} \vec 1'\Sigma^{-1}\ve  \alpha }\right) \frac{\Sigma^{-1}\vec 1}{\vec 1'\Sigma^{-1}\vec 1} + \frac{1}{\gamma}\Sigma^{-1}\ve  \alpha. \nonumber
\end{align}
using $A = \vec 1' \Sigma^{-1}\vec 1$ and $B = \ve \alpha' \Sigma^{-1}\vec 1 = \vec 1'\Sigma^{-1}\ve \alpha$ we find
\begin{equation}\label{eq:mv4}
      \ve \theta^{*}_7 = \left({g_0 - \frac{1}{\gamma} B}\right) \frac{\Sigma^{-1}\vec 1}{A} + \frac{1}{\gamma} {\Sigma^{-1}\ve \alpha}.
\end{equation}
The global minimum variance portfolio is $\ve \theta_0 = \Sigma^{-1}\vec 1/A$. The sum of its weights are unity: $\vec 1' \ve \theta_0=1$. The optimal portfolio can be written as the weighted sum of the GMV portfolio and the risky $\ve \alpha$ portfolio $\ve \theta_{\alpha}=\Sigma^{-1} \ve \alpha/B$:
\begin{equation} \label{eq:mvp}
    \ve \theta^*_7 = \left({g_0 - \left[ {\frac{1}{\gamma} B}\right]}\right) \ve \theta_{0} + \left[ {\frac{1}{\gamma}B}\right] \ve \theta_{\alpha}.
\end{equation}
Here with $\mu = \sfrac{1}{\gamma}B$ we have: ${\ve \theta_7}^* = (g_0- \mu) \ve \theta_0 + \mu \ve \theta_{\alpha}$.

We find that $\ve \theta^*_7 = g_0 \ve \theta_0 - \mu (\ve \theta_0 - \ve \theta_{\alpha})$, which is a combination of the global minimum risk portfolio and an optimal risky cash neutral long/short portfolio: $\vec 1' (\ve \theta_0 - \ve \theta_{\alpha}) = 0$. We can factor out the global minimum variance portfolio $\ve \theta_0$ in Eqn.(\ref{eq:mv4}) and with a bit of matrix algebra we see that the portfolio is made of a geared global minimum variance portfolio, whose weights sum to one, and a risky portfolio which is dependent on the return expectations and risk aversion \cite{lee2000theory} but can have gearing:
\begin{equation} \label{eq:mvgearing6}
    \ve \theta^*_7 = \left[{ g_0  + \frac{1}{\gamma} \Sigma^{-1}(\ve \alpha \vec 1' - \vec 1 \ve \alpha')}\right] \ve \theta_0. % \frac{ \Sigma^{-1} \vec 1}{A}.
\end{equation}
By considering the $i$-th and $j$-th components on the inner bracket in the matrix product in the second  term, it can be seen to be driven by the pairs of relative $\ve \alpha$: $(\ve \alpha \vec 1' - \vec 1 \ve \alpha')_{ij} = \ve \alpha_i - \ve \alpha_j$ where $\diag(\ve \alpha \vec 1' - \vec 1 \ve \alpha')=0$. The first term on the left in the bracket is the geared global minimum variance portfolio where, the weights of the GMV portfolio sum to one, but is now multiplied by the gearing parameter $g_0$. The second term is a portfolio whose weights sum to zero, but it is leveraged by risk aversion and then projected onto the global minimum risk portfolio. The risky portfolio is independent of the gearing parameter $g_0$.

We can contrast Eqn.(\ref{eq:mvp}) with Eqn.(\ref{eq:minriskgear5}); and set $g_0=1$ to find that $\ve \theta^*_6 = \ve \theta_0$ for the risk minimisation. We see that with $g_0=1$ we have recovered the Global Minimum Variance (GMV) portfolio $\ve \theta_0$; as in Eqn.(\ref{eq:mvgearing6}) when we let the risk aversion become sufficiently large. Both portfolio optimisations have the same minimum variance portfolio, as expected following from a two fund separation, and the minimum variance portfolio are combined with a risky portfolio. This is contrasted with the unconstrained optimisations given in Table (\ref{tab:unconstrainedMV}) which are all leveraged versions of the optimal risky portfolio $\ve \theta_{\alpha}$.

Finally, we have that from Eqn.(\ref{eq:mvp}) that the optimal portfolio is not generally aligned with $\ve \theta_{\alpha}$ and hence we cannot explain the lower bound deviation from alignment with $\ve \alpha$ being due to the condition number. However, we can show that there is a weakened condition number lower bound. Eqn.(\ref{eq:mv4}) can be written in the form of $\Sigma^{-1} \vec w$
\begin{equation} \label{eq:gearingw}
   \ve \theta^*_7 = \frac{1}{\gamma} \Sigma^{-1} \vec w ~\mbox{where}~ \vec w = \left[ { \left({g_0 - \frac{1}{\gamma} B}\right) \frac{\vec 1}{A} + \frac{1}{\gamma} \ve \alpha} \right],
\end{equation}
where $\vec w$ is the same dimension as $\ve \alpha$. By Theorem \ref{thm:bah}, as for the risk minimisation with a gearing constraint in Eqn.(\ref{eq:minriskgear5}), we again know that the angle between this portfolio and the $\ve \alpha$ portfolio has a upper bound dependent on the condition number of the estimated covariance matrix. It is still advantageous to tilt the gearing constrained optimal risky portfolio within the overall optimal portfolio towards $\ve \alpha$ by improving the conditioning of the covariance matrix, while maintaining the constraint. However, in this instance the gearing can saturate any information advantage as the gearing can drag the portfolio away from the direction of $\ve \alpha$ towards the global minimum variance portfolio, even if in the risky portfolio, the angle can be reduced by improving the conditioning. 

\subsection{Geometric interpretation}

%%Discuss how the extended result can be used and why it is interesting.
The key geometric insight of \cite{goltsandjones2009} is to consider the expected portfolio return as a function of the angle $\varphi$: $\alpha_P = \ve \alpha' \ve \theta = |\ve \alpha | |\ve \theta| \cos{\varphi}$. This splits the problem into a direction and magnitude and shows that to ameliorate the impact of poor covariance conditioning it is sufficient to constrain the angle $\varphi$. Here the investment ``direction'' is given by the unit vector $\ve{ \hat \theta} = \ve \theta / |\ve \theta|$ associated with the optimal portfolio weight in the portfolio weight space -- which will be some simplex. This is visualised in Figure \ref{fig:geometry1} for the optimisations I, II, III, IV, V and VIII. The gearing constrained optimisations are visualised in Figure \ref{fig:geometry2}. The investment ``magnitude'' is given by the $|\ve \theta|$. 

The expected portfolio return can be re-written using either of the portfolio solutions (either $\Sigma^{-1} \vec z$ from Eqn. (\ref{eq:minriskgear5}) or $\Sigma^{-1} \vec w$ from Eqn. (\ref{eq:mvp})). Both are of the form: $g_0 \ve \theta_0 - w (\ve \theta_0 - \ve \theta_{\alpha})$, for some weight $w \in [\mu, \omega]$: $\alpha_P = \ve \alpha' \ve \theta = (g_0-w) \ve \alpha' \ve \theta_0 + w \ve \alpha' \ve \theta_{\alpha}$. This has two angles, the first between $\ve \alpha$ and the global minimum variance portfolio, some angle $\varphi_0$, and the second between the optimal risk portfolio and $\ve \alpha$, $\varphi_1$. The overall $\ve \alpha$-angle $\varphi$ can be decomposed into these two angles:
\begin{equation} \label{eq:varphi01}
    \cos(\varphi) = (g_0 - w) \frac{|\ve \theta_0|}{|\ve \theta|} \cos \varphi_0 + w \frac{|\ve \theta_{\alpha}|}{|\ve \theta|} \cos \varphi_1.
\end{equation}
If we improve the conditioning of the covariance matrix we would be able to reduce the upper bound on the second angle $\varphi_1$ (by definition $\phi$), the angle between the optimal risky portfolio and $\ve \alpha$. This would in turn reduce the overall angle between the optimal portfolio and the direction of $\ve \alpha$. However, improving the conditioning of the covariance matrix does not mean the $\varphi_0$ will lead to the overall portfolio being closer to $\ve \alpha$. It may well drag the optimal portfolio away from the expected $\ve \alpha$ because it could move the minimum variance portfolio away from the expected returns. This suggests that one should really be minimising $\varphi_1$ {\it i.e.} minimise the angle between the optimal risky portfolio and $\ve \alpha$ (the unconstrained portfolio optimisation), rather than the overall $\ve \alpha$ angle itself (the full constrained portfolio optimisation).

From the first order conditions\footnote{${\partial_\varphi} \cos \varphi = 0$} and Eqn. (\ref{eq:varphi01}) : 
\begin{equation} \label{eq:fvarphi1}
        \frac{\partial \varphi_0}{\partial \varphi_1} = -\frac{w}{(g_0-w)} \frac{\sin \varphi_1}{\sin \varphi_0} \frac{|\ve \theta_{\alpha}|}{|\ve \theta_0|}.
\end{equation}
Minimising with respect to the overall angle $\varphi$ is equivalent to minimising with respect to $\varphi_1$ because from Eqn.(\ref{eq:varphi01}) and integrating Eqn.(\ref{eq:fvarphi1}) where there is some well behaved function $f$ such that $\varphi_0 = f(\varphi_1)$ on $\varphi_1 \in [0,\pi/2]$. If we minimise the $\ve \alpha$-angle $\varphi$ we would have minimised $\varphi_1$ (equivalently $\phi$), the angle between the optimal risky portfolio $\ve \theta_{\alpha}$ and the direction of the expected returns $\ve \alpha$.  It is sufficient to minimise the angle between the optimal risky portfolio and the direction of the expected returns $\varphi_1=\phi$ in order that the overall $\ve \alpha$-angle is minimised in order to achieve optimal outcomes even in the presence of the gearing constraint. We now consider precisely how to achieve such an optimal outcome when there is estimation uncertainty without making strong distributional assumptions. 

\section{Robust portfolio optimisations}

\subsection{Constraining the $\ve \alpha$-angle}

\citet{goltsandjones2009} argue that the mean-variance optimisations can be made robust by constraining the $\ve \alpha$-angle. They considered an uncertainty region $U_{\alpha}$ as a ball centred on $\ve \alpha$ with radius:
\begin{equation}
    r_{\alpha} = k | \ve \alpha | ~\mbox{for}~ k \in (0,1).
\end{equation}
This is some 1-$\sigma$ neighbourhood. The expected portfolio return in this neighbourhood is: $\alpha_p = \ve \theta'  \ve \alpha$, and the expected return of the optimal risky portfolio in the same neighbourhood is: $\alpha_r = \ve \theta_{\alpha}' \ve \alpha$. In the unconstrained cases given in Table \ref{tab:unconstrainedMV} the robust expected returns of the optimal risky portfolios
\begin{equation} \label{eq:riskyretmin1}
    \min_{U_{\alpha}} \alpha_r = \ve \theta_{\alpha}' \ve \alpha - k |\ve \alpha||\ve \theta_{\alpha}|,
\end{equation}
are equivalent to the robust expected returns of the optimal portfolio  \citep{goltsandjones2009}:
\begin{equation} \label{eq:riskyretmin2}
    \min_{U_{\alpha}} \alpha_p = \ve \theta' \ve \alpha - k |\ve \alpha||\ve \theta|.
\end{equation}
Here $\alpha_p=\alpha_r$. This is not in general the case in the presence of the gearing constraint, in particular for optimisations VI and VII in Table \ref{tab:linearcons}, because $\ve \theta_{\alpha} \neq \ve \theta^*$ and hence $\alpha_r \neq \alpha_p$. 

Our objective is to minimise the angle between the optimal risky portfolio and the expected return in the case of the gearing constraint in order to ameliorate the impact of poor conditioning of the covariance, however, this is equivalent to minimising the overall angle between the optimal portfolio and the expected returns -- see Eqn.(\ref{eq:varphi01}). This implies that from expected portfolio return we can minimise this over the $\ve \alpha$-weight angle $\varphi$:
\begin{equation} \label{eq:Uball}
    \min_{U_{\alpha}} \alpha_p = |\ve \theta| |\ve \alpha| (\cos(\varphi) - k) ~ \mbox{for}~ k \in (0,1).
\end{equation}
The optimizations now include the expected returns but with a regularisation term.

In practice this can be achieved by minimising the angle between the optimal risky portfolio $\ve \theta_{\alpha}$ and the expected returns $\ve \alpha$ by shrinking the covariance matrix because the difference in the direction of the optimal risky portfolio and $\ve \alpha$ is entirely due to the covariance $\Sigma$: $\ve \theta_{\alpha} \propto \Sigma^{-1} \ve \alpha$. Hence we try find some new covariance $\tilde \Sigma$ that will approximate the required $\ve \alpha$-angle minimisation. This now links the method of \cite{goltsandjones2009} to conventional shrinkage methods that focus entirely, and exclusively, on the conditioning of the covariance matrix.

The reformulated portfolio optimizations are then as in Table (\ref{tab:robust}). What is important to realise is that the less one trusts the expected asset returns $\ve \alpha$ the larger we make the regularisation hyper-parameter $k$. This ensures that the less certain we are about the asset returns, the closer to the expect returns the risky portfolio will be, and in turn the closer our portfolio will be -- this ensures that we do not excessively leverage the errors in the expected asset returns. The key difference in the presence of the constraint is that we will optimally mix in the equally weighted portfolio so that the overall optimal portfolio is not always aligned with the optimal risky portfolio.

\begin{table}[ht]% \sidecaption
%\scriptsize
\centering
\begin{tabular}{p{1cm}lrl}
\toprule
\multicolumn{3}{c}{$\ve \alpha$-angle Robust Portfolio Choice} \\
\midrule
\# & Optimisation & Constraints \\
\midrule
I, V & $\arg\displaystyle\max_{\ve \theta}  \left\{ {\displaystyle\min_{U_{\alpha}} \alpha_p} \right\}$ & \multicolumn{2}{r}{$\sigma^2_p=\ve \theta' \Sigma \ve \theta \le \sigma^2_0$} \\ 
II & $\arg\displaystyle\min_{\ve \theta} \left\{ { \frac{1}{2} \ve \theta' \Sigma \ve \theta} \right\}$ & \multicolumn{2}{r}{$\displaystyle\min_{U_{\alpha}} \alpha_p = \ve \theta' \ve \alpha \ge \alpha_0$} \\ 
III & \multicolumn{2}{l}{$\arg\displaystyle\max_{\ve \theta}  \left\{ {\displaystyle\min_{U_{\alpha}} \alpha_p - \frac{\gamma}{2} \ve \theta' \Sigma \ve \theta} \right\}$} & \\ 
IV,VIII & $\arg\displaystyle\max_{\ve \theta}  \left\{ \frac{  \displaystyle\min_{U_{\alpha}} \alpha_p}{\ve \theta' \Sigma \ve \theta}  \right\}$ & &  \\
\toprule
VI & $\arg\displaystyle\min_{\ve\theta} \Bigg\{ {\frac{1}{2}\ve \theta'\Sigma\ve \theta} \Bigg\}$ & \multicolumn{2}{r}{$\begin{matrix}\displaystyle\min_{U_{\alpha}} \alpha_p = \ve \alpha' \ve \theta \ge \alpha_0 \\  \vec 1' \ve \theta = g_0\end{matrix}$}\\
VII & \multicolumn{2}{l}{$\arg\displaystyle\max_{\ve \theta} \left\{ { \displaystyle\min_{U_{\alpha}}\alpha_p - \cfrac{\gamma}{2} \ve \theta'\Sigma \ve \theta }\right\}$} & $\vec 1' \ve \theta = g_0$ \\
\bottomrule
\end{tabular}
\caption{Two linearly constrained portfolio optimisation problems, VI and VII, can then be extend into the robust geometric framework of \cite{goltsandjones2009} (See Table \ref{tab:unconstrainedMV}) by including a minimisation over the ball $U_{\alpha}$ defined in Eqn.(\ref{eq:Uball}). Optimization V from Table \ref{tab:linearcons} is equivalent to optimization I because fixing $\sigma_p$ fixes the gearing $g_0$. Optimization IV is equivalent to VIII where the solution of IV is scaled by $g_0$.} \label{tab:robust}
\end{table}

\subsection{Shrinking with the $\ve \alpha$-angle} \label{ssec:alphashrinking}

Estimating covariance matrices of stock returns has always been a problematic area. The sample covariance unfortunately creates a well documented problem, where if the number of stocks under consideration are large, particularly relative to the number of observations, the sample covariance matrix contains a lot of error \citep{jak1980}. The most extreme coefficients in the matrix tend to take on extreme values not because they are correct, but because they contain an extreme amount of error. This also implies that the condition number of covariance matrices tend to be large, and very sensitive to the input data. Covariance shrinkage can ameliorate this, as shown by \citet{ledoitwolf2004}. Here we argue that the shrinkage itself should be considered in the context of the optimisation constraints.

Following \cite{goltsandjones2009} one can consider a shrunk covariance for a portfolio with $n$ assets. For example, we can consider the Sharpe ratio maximization solution, optimization IV, but with a new covariance where $k \in [0, 1]$:
\begin{equation} \label{eq:cov2}
    \tilde \Sigma = \frac{k|\ve \alpha| |\ve \theta|}{\sqrt{\ve \theta' \Sigma \ve \theta}} I + \left({ \frac{\ve \alpha' \ve \theta -  k|\ve \alpha| |\ve \theta|}{\sqrt{\ve \theta' \Sigma \ve \theta}}}\right) \Sigma.
\end{equation}
Here $\Sigma$ is the original estimated covariance matrix, $I$ is the identity matrix, $\ve \theta$, $\ve \alpha$ and $k$ are as before. When $k$ is zero we recover the optimal risky portfolio $\ve \theta_{\alpha} = \Sigma^{-1} \ve \alpha/B$ by solving: $\ve \theta = \tilde \Sigma^{-1} \ve \alpha$ to recover the Sharpe ratio maximisation solution of the optimal portfolio $\ve \theta^*$ for optimization IV. This is just the optimal risky portfolio $\ve \theta_{\alpha}$. This construction is valid for optimisations I, II, III, VI, V and VIII as these are all scalar multiples of Sharpe ratio maximisation optimal risky portfolio. 

We can eliminate a direct dependency on the optimal portfolio returns and risk and set the optimal risky portfolio to be fully invested ($g_0=1$), so that from Eqn.(\ref{eq:cov2}) with $k \in[0,\cos \varphi]$ and using that $\ve \alpha' \ve \theta=|\ve \alpha| |\ve \theta| \cos({\varphi})$ we can pick a covariance that updates the optimal solution directly to find $\tilde {\ve \theta}_{\alpha} = \tilde \Sigma^{-1} \ve \alpha/{\tilde B}$ where:
\begin{equation} \label{eq:shrinkcov}
     \tilde \Sigma = \frac{k}{\cos \varphi} I + \left({1 -  \frac{k}{\cos \varphi}}\right) \Sigma.
\end{equation}
Here $k$ can be understood as some quantity $\cos (\varphi_k)$ for some targeted optimal angle $\varphi_k \in[\varphi,\pi/2]$ between the risky portfolio and the expected returns. When $\Sigma \to I$ will recover the same limiting solution as $k \to \cos \varphi$ and align the optimal portfolio with the direction of the expected returns: $\hat {\ve \alpha}$. Any of the unconstrained optimal portfolio solutions can be modified in this way using an appropriate chosen covariance that would shrink the covariance towards the identity matrix so that the portfolio can be shifted between the original solution and a regularised one. This is equivalent to conventional covariance shrinkage methods.

We can now extend this to find a new regularised covariance given a prudent choice of $k$ from Eqn.(\ref{eq:shrinkcov}) for any of the gearing constrained optimisations. We compute a new global minimum variance portfolio $\ve {\tilde \theta_0}$ and a new optimal risky portfolio $\tilde {\ve \theta}_{\alpha}$ with $k \in [0,\cos \varphi]$. Then, we combined these to form a new efficient frontier exploiting the two fund separation theorems. This ensures that we have retained consistency with the constraints, while minimising the $\ve \alpha$-angle as in Eqn.(\ref{eq:Uball}). Optimisations VI and VII are combinations of the GMV portfolio and the optimal risky portfolio. The overall portfolio is built up of the combination of two portfolios: one with direction $\tilde \Sigma^{-1} \vec 1$, and the other with the direction of $\tilde \Sigma^{-1} {\ve \alpha}$. When one is less certain about the asset returns $\ve \alpha$ then the regularisation parameter $k$ is chosen to be larger. The choice of any hyper-parameter such as the regularisation parameter requires some sort of cross-validation. Hence in practice it should be expected that one would need to engage in simulation work to best decided on the value of k in a way that was not overly prone to backtest over-fitting \citep{bailey2015}.

In the presence of the gearing constraint, the greater $k$, the more exposed we are to the $1/n$ portfolio factored from the GMV portfolio. This is because when $k$ increases towards $k_0=\cos \varphi$ we will shrink the covariance towards the identity matrix $I_n$. This will shrink the GMV portfolio towards the equally weighted $1/n$ portfolio given by a unit vector $\hat {\vec e}$:
\begin{equation}
   \ve {\tilde \theta}_0 = \frac{\tilde \Sigma^{-1} \vec 1}{\vec 1' \tilde \Sigma^{-1} \vec 1} \xrightarrow{k \to k_0} \frac{\vec 1}{\vec 1' I_n \vec1} =\frac{\vec 1}{n} = \hat {\vec e}.
\end{equation}
Here $\hat {\vec e}' \vec 1 =1$, where $\vec 1' \vec 1= n$. The risky portfolio becomes the weighted average of the expected returns which is just the unit vector in the direction of $\ve \alpha$: 
\begin{equation}
   \ve {\tilde \theta}_{\alpha} = \frac{\tilde \Sigma^{-1} \ve \alpha}{\vec 1' \tilde \Sigma^{-1} \ve \alpha} \xrightarrow{k \to k_0} \frac{I_n \ve \alpha}{\vec 1' I_n \ve \alpha} = \frac{\ve \alpha}{\ve \alpha' \vec 1}= \hat {\ve \alpha}.
\end{equation}
Here $\hat {\ve \alpha}' \vec 1 = 1$ and $\ve \alpha' \vec 1 = n \bar \alpha = |\ve \alpha|$ for the average alpha $\bar \alpha$.
The combination of the GMV portfolio, $\tilde {\ve \theta}_{0}$, and the optimal risky portfolio, $\tilde  {\ve \theta}_{\alpha}$, is controlled by the amount of gearing $g_0$, as seen in equations (\ref{eq:minriskgear5}) and (\ref{eq:mvp}). The combination remains cash neutral: $\vec 1' (\tilde {\ve \theta_0} - \tilde {\ve \theta_{\alpha}}) = 0$. 

For the gearing constrained mean-variance optimisation given by Eqn.(\ref{eq:mvp}) and using covariance shrinkage towards the identity matrix will have $\tilde A=n$ and $\tilde B = \ve \alpha' \vec 1$ to find from Eqn.(\ref{eq:mvgearing6}):
\begin{equation}
    {\tilde {\ve \theta}_6^*}  = g_0 \hat {\vec e} + \frac{1}{\gamma} I_n \left( {\ve \alpha \vec 1' - \vec 1 \ve \alpha'} \right) \unitv.
\end{equation}
The average expected return is: $\bar \alpha = \frac{1}{n} \ve \alpha' \vec 1$ and this can be used to find: 
\begin{equation} \label{eq:mvgearingS1}
     {\tilde {\ve \theta}_6^*}  = g_0 \unitv + \frac{1}{\gamma} \left( {\ve \alpha - \vec 1 \bar \alpha} \right).
\end{equation}
We can confirm this by checking that the last term correctly sums to zero when multiplied by $\vec 1'$. From the unit vector in the direction of $\ve \alpha$: $|\ve \alpha| = n \bar \alpha$ where $\ve \alpha = |\ve \alpha|\hat{\ve \alpha}$ and using this we can write Eqn.(\ref{eq:mvgearingS1}) as:
\begin{equation}
       {\tilde {\ve \theta}}^*_6  = g_0 \unitv + \frac{|\ve \alpha|}{\gamma} \left( {\hat {\ve \alpha} - \unitv} \right).
\end{equation}
This is geometrically intuitive because we see that when the $\ve \alpha$ direction $\hata$ aligns with the equally weighted portfolio $\unitv$ then our best case optimal portfolio is just the geared GMV and independent of $\ve \alpha$ because of the constraint.   

For the case of the risk minimisation the $\ve \alpha$ standard deviation will be required: $n^2 \sigma^2_{\alpha} = n \ve \alpha' \ve \alpha - (\ve \alpha' \vec 1)^2$. From Eqn.(\ref{eq:minriskgear5}) again with $\tilde A=n$, $\tilde B = \ve \alpha' \vec 1$, $\tilde C = \ve \alpha' \ve \alpha$ and $\tilde D = n \ve \alpha' \ve \alpha - (\ve \alpha' \vec 1)^2$. We can identify $\tilde D$ with the $\ve \alpha$ variance: $\tilde D = n^2 \sigma_{\alpha}^2$. For optimization VII:
\begin{equation}
    {\tilde {\ve \theta}}^*_7= g_0 \unitv + \tilde \omega (\hat{\ve \alpha}-\unitv).
\end{equation}
Here the weight is: $\tilde \omega = (g_0 (\ve \alpha' \vec 1)^2- n \alpha_0 (\ve \alpha' \vec 1))/\tilde D$. Putting this all together we find that the maximally shrunk optimal risk minimisation with gearing constraint is:
\begin{equation}
     {\tilde {\ve \theta}}^*_7= g_0 \unitv + \left[{\frac{g_0 (\ve \alpha' \vec 1)^2- n \alpha_0 (\ve \alpha' \vec 1)}{n \ve \alpha' \ve \alpha - (\ve \alpha' \vec 1)^2}}\right] (\hat{\ve \alpha}-\unitv).
\end{equation}
This can then be written as:
\begin{equation}
   {\tilde {\ve \theta}}^*_7=g_0 \unitv + \frac{1}{\sigma^2_{\alpha}} \left[{g_0 (\ve \alpha' \unitv)^2- \alpha_0 (\ve \alpha' \unitv)}\right] (\hat{\ve \alpha}-\unitv).
\end{equation}
From the definition of the average expected returns:
\begin{equation} \label{eq:uncertain1}
    {\tilde {\ve \theta}}^*_7= g_0 \unitv + \frac{1}{\sigma^2_{\alpha}} \left({g_0 \bar \alpha^2 - \alpha_0 \bar \alpha}\right) (\hat{\ve \alpha}-\unitv).
\end{equation}
When the $\ve \alpha$ aligns with the equally weighted portfolio we find the best case portfolio is once again the geared equally weighted portfolio: $g_0 \unitv$. When the $\ve \alpha$ is orthogonal to the equally weight portfolio $\ve \alpha' \unitv =0$ we again find some combination. 

In the unconstrained optimisation the risky portfolio is shrunk towards the asset views $\ve \alpha$; this is no longer the case in the presence of gearing. There will be a combination of an equally weighted portfolio and $\ve \alpha$. However, the gearing contributions can be conveniently factored out and treated separately relative to the unconstrained solution. 

%tikz
\begin{figure}[ht!]
    \centering
    %\resizebox{\columnwidth}{!}{
        \tdplotsetmaincoords{25}{110}
        \begin{tikzpicture}[scale=3,tdplot_main_coords]
            
            \coordinate (O) at (0,0,0);
            %coordinate system
            \draw[thick,->] (-1,0,0) -- (1,0,0) node[anchor=north east]{$\omega_1$};
            \draw[thick,->] (0,-1,0) -- (0,1,0) node[anchor=north west]{$\omega_2$};
            \draw[thick,->] (0,0,-0.5) -- (0,0,1) node[anchor=south]{$\omega_3$};
            
            % alpha vector
            \tdplotsetcoord{A}{0.9}{50}{240};
            %draw a vector from origin to point (P)
            \draw[very thick, -stealth,color=red] (O) -- (A) node[anchor=south]{$\ve \alpha$};
            %draw projection on xy plane, and a connecting line
            \draw[dashed, color=red] (O) -- (Axy);
            \draw[dashed, color=red] (A) -- (Axy);
            
             % optimal risky portfolio
            \tdplotsetcoord{R}{1.3}{50}{250};
            %draw a vector from origin to point (P)
            \draw[-stealth,color=red] (O) -- (R) node[anchor=south]{$\ve \theta_{\alpha}$};
            %draw projection on xy plane, and a connecting line
            \draw[dashed, color=red] (O) -- (Rxy);
            \draw[dashed, color=red] (R) -- (Rxy);
            
              % optimal portfolio
            \tdplotsetcoord{T}{1.1}{40}{320};
            %draw a vector from origin to point (P)
            \draw[very thick,-stealth,color=black] (O) -- (T) node[anchor=east]{$\tilde{\ve \theta}^*$};
            %draw projection on xy plane, and a connecting line
            \draw[dashed, color=black] (O) -- (Txy);
            \draw[dashed, color=black] (T) -- (Txy);
            
              % optimal portfolio
            \tdplotsetcoord{S}{0.7}{45}{340};
            %draw a vector from origin to point (P)
            \draw[very thick,-stealth,color=black] (O) -- (S) node[anchor=east]{${\ve \theta}^*$};
            %draw projection on xy plane, and a connecting line
            \draw[dashed, color=black] (O) -- (Sxy);
            \draw[dashed, color=black] (S) -- (Sxy);
            
            % global minimum variance portfolio
            \tdplotsetcoord{G}{1}{60}{10};
            %draw a vector from origin to point (P)
            \draw[-stealth,color=blue] (O) -- (G) node[anchor=west]{$\ve \theta_0$};
            %draw projection on xy plane, and a connecting line
            \draw[dashed, color=blue] (O) -- (Gxy);
            \draw[dashed, color=blue] (G) -- (Gxy);
            
            % Shrunk portfolios
              % optimal risky portfolio
            \tdplotsetcoord{E}{0.7}{45}{45};
            %draw a vector from origin to point (P)
            \draw[very thick,-stealth,color=blue] (O) -- (E) node[anchor=west]{$\unitv$};
            %draw projection on xy plane, and a connecting line
            \draw[dashed,color=blue] (O) -- (Exy);
            \draw[dashed,color=blue] (E) -- (Exy);
            
            % draw the arc
            \tdplotdefinepoints(0,0,0)(1.8,-0.6,2)(-1.1,-2,2);
            \tdplotdrawpolytopearc[thick,color=black,stealth-stealth]{0.5}{anchor=north west}{$\varphi$};
            
            % draw the arc
            \tdplotdefinepoints(0,0,0)(-1.2,-2,1)(-1.1,-1.5,1);
            \tdplotdrawpolytopearc[color=red, -stealth]{1.2}{anchor=north west}{\textcolor{red}{$k \to k_0$}};
            
            % draw the arc
            \tdplotdefinepoints(0,0,0)(3,0.5,1)(0.33,0.33,0.33);
            \tdplotdrawpolytopearc[color=blue,-stealth]{1}{anchor=north west}{\textcolor{blue}{$k \to k_0$}};
            
             % draw the arc
            \tdplotdefinepoints(0,0,0)(0.7,-0.4,0.5)(0.8,-1,0.7);
            \tdplotdrawpolytopearc[color=black,-stealth]{0.8}{anchor=north east}{$k \to k_0$};
        \end{tikzpicture}
    %}
    \caption{The stylised geometry of the robust portfolio construction for optimisations VI and VII. Here $\unitv$ is the unit vector $\unitv' \vec 1 =1$ and $\hat {\ve \alpha}$ is the direction of the expected returns. The equally weighted portfolio is given by $\unitv$. With the shrinkage of the covariance, $\Sigma \to \tilde \Sigma$ as $k \to k_0$, the optimal portfolio $\ve \theta^*$ is rotated to $\tilde{\ve \theta}^*$. For optimisations VI and VII this will be some linear combination of $\hat {\ve \alpha}$ and $\unitv$. The optimal risky portfolio $\ve \theta_{\alpha}$ is rotated towards $\ve \alpha$, and the global minimum variance portfolio is rotated towards $\unitv$.  Optimisations I, II, III, IV, V and VIII rotate towards $\ve \alpha$.}
    \label{fig:geometry3}
\end{figure}

\section{Simulated Portfolios} \label{sec:simports}

\subsection{Simulated Pareto surfaces} \label{ssec:efffront}

\begin{figure*}[h!]
    \centering
    \subfloat[][No explicit gearing constraint with $g_0=1$]{
       \includegraphics[width=0.5\textwidth]{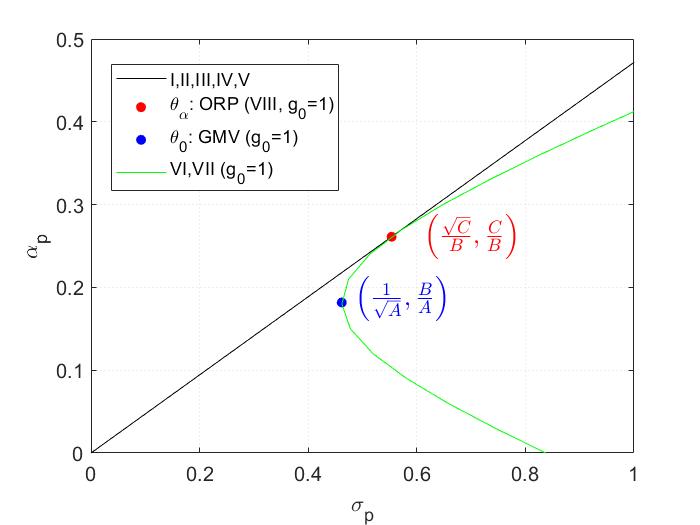}
    }
    \subfloat[][Gearing constrained $g_0 \in (0.5,1.5)$]{
       \includegraphics[width=0.5\textwidth]{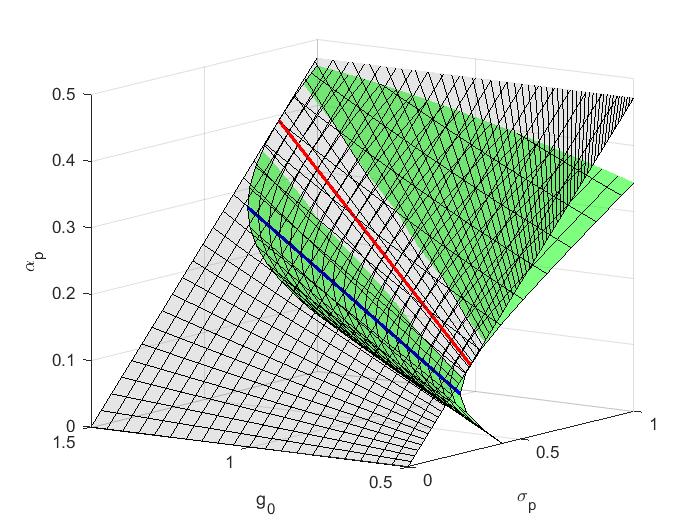}
    }
    \caption{Fig (a) is the Pareto frontier sliced from Fig (b) at $g_0=1$. This visualises and compare the stylised Pareto (efficient) frontiers for the optimizations problems I, II, III and VI from Table \ref{tab:unconstrainedMV} with those that have explicit gearing constraints: optimisations V, VI, VII and VIII from Table \ref{tab:linearcons}. Optimisations I,II, III and IV are portfolios that generate a Capital Market Line (CML) as they can be considered linear combinations of some risk-less portfolio $\vec 0$ and the optimal risky portfolio $\ve \theta_{\alpha}$.}
    \label{fig:efffront}
\end{figure*}

The efficient frontiers for the different optimisations are shown in Figure \ref{fig:efffront}. Optimisations I, II, III and VI from Table \ref{tab:unconstrainedMV} are shown in Figure \ref{fig:efffront}(a). These are implicitly geared versions of the fully invested optimal risky portfolio and hence generate a straight line in the $(\alpha_p,\sigma_p)$ space that can be identified as the Capital Market Line (CML) associated with a 0\% risk-free rate. The fully invested version of optimisations VI and VII from Table \ref{tab:linearcons} are shown along with the GMV portfolio. Figure \ref{fig:efffront}(b) depicts a stylised version of optimisations V, VI, VII and VIII from Table \ref{tab:linearcons}. Optimisations V and VIII generate the Capital Market surface (the grey plane) which intersects the solutions to optimisations VI and VII (the green curved surface) along the line representing the family of geared optimal risky portfolios (the red line). The family of geared global minimum risk portfolios is also shown (the blue line). The figures are indicative and demonstrate the well known general geometry of the fully invested portfolio optimisation Pareto surfaces. Figure \ref{fig:efffront}(a) is the slice at $g_0=1$ from the surface shown in Figure \ref{fig:efffront}(b). 

%% large surface figure (sigma_p.alpha_p,g_0) 

\subsection{Simulated $\alpha$-angle dependency} \label{ssec:alphadepend}

The equally-weighted portfolio has been a benchmark in the portfolio selection literature and we consider what the $\alpha$-weight angle of equally-weighted portfolios are compared to those of portfolios resulting from mean variance optimisation. Here we demonstrate the relative performance relation between the equally weighted portfolio and the mean variance optimal portfolio via direct simulation. This relationship was investigated by simulating 10 years worth of monthly stock returns, for 10 stock portfolios, from normal distributions with random mean (between 0\% and 50\% p.a) and random standard deviation (between 0\% and 20\% p.a.) with 1000 replications per simulation. 

\begin{figure*}[h!]
\centering 
    \subfloat[][Unconstrained performance (I)]{
       \includegraphics[width=0.5\textwidth]{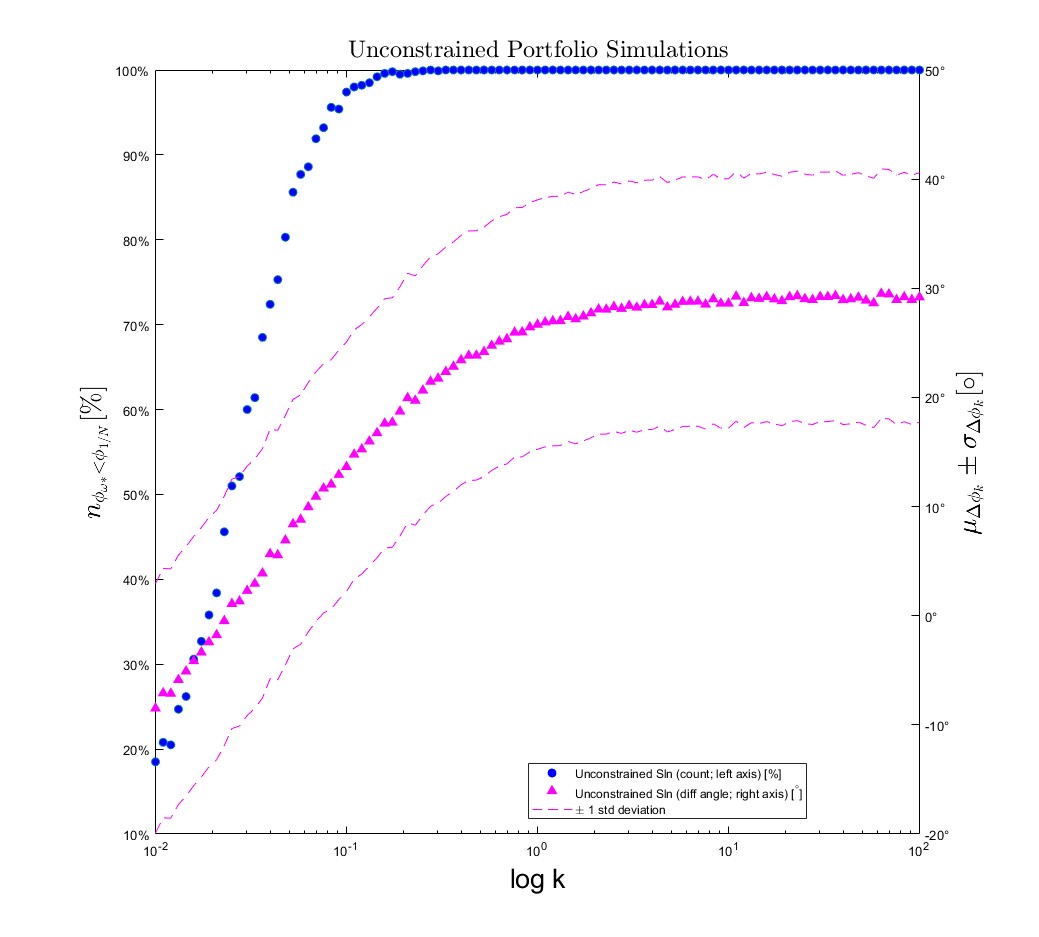}
    }
    \subfloat[][Constrained performance (VI \& VII)]{
       \includegraphics[width=0.5\textwidth]{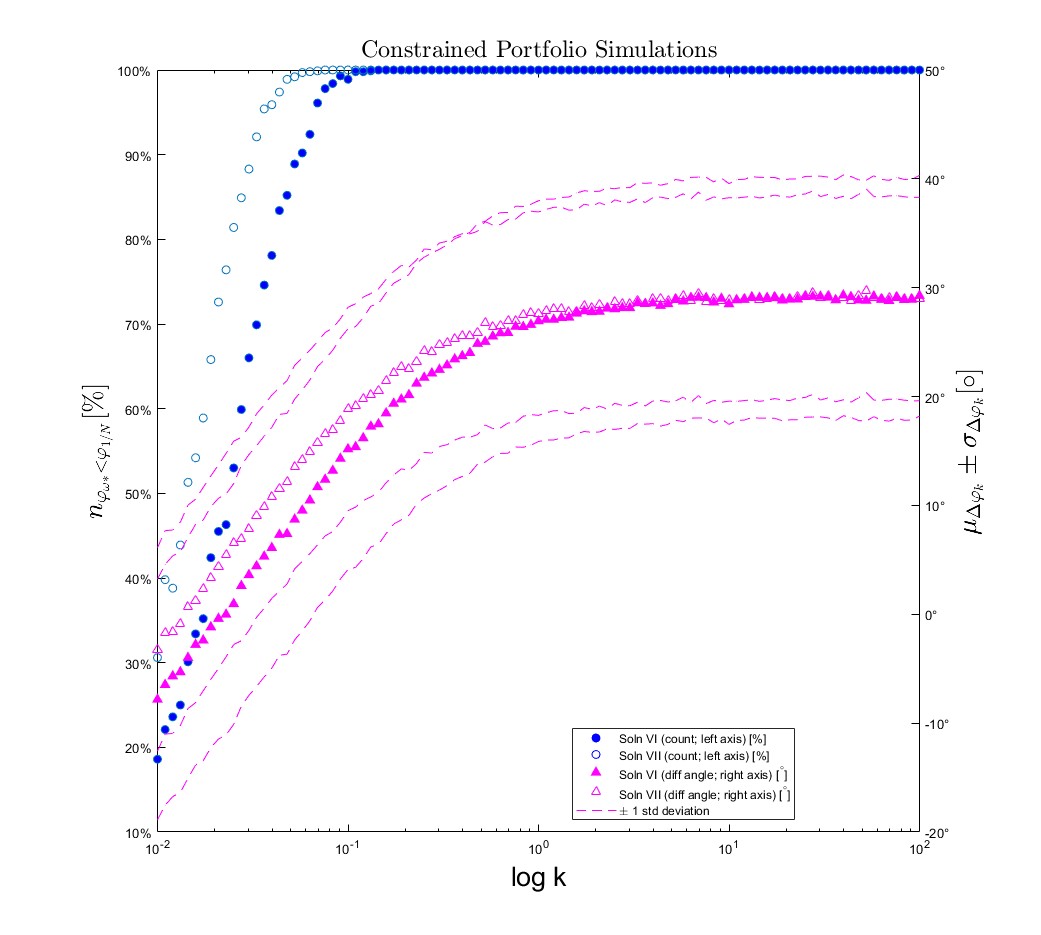}
    } 
         \caption{Both figures show the percentage number $n_{\phi_{\omega^*}<\phi_{1/N}}$ of performance simulations with weights better aligned with those of the returns (in circles against the left axis), and the average difference (in triangles against the right axis) between the conditioned angles and the unconditioned angles $\mu_{\Delta \phi_k} = \E_m[\phi_k-\phi_0]$ with confidence intervals. Fig. \ref{fig:normal_distr_random_sims_AW}(a) is for the unconstrained optimisations, and Fig. \ref{fig:normal_distr_random_sims_AW}(b) for the constrained cases (VI and VII). Each point represents $m=1000$ replications. This shows the unconditioned optimisations ($k$ =0) relative to those with \citet{goltsandjones2009} covariance conditioning ($k> 0$). 100\% of the simulations show smaller $\alpha$-weight angles for $k> 10^{-1}$ on the resulting optimal risk portfolios relative to the the equally weighted portfolio. On average, this angle is just under 30$^\circ$ better. \cite{goltsandjones2009} conditioned optimal portfolios outperform the equally weighted portfolios.}
       \label{fig:normal_distr_random_sims_AW}
\end{figure*}

First, the $\alpha$-weight angle for the equally weighted portfolio was compared to the $\alpha$-weight angle of the optimal risk portfolio ($\theta_\alpha$) on every simulation.  Since all unconstrained mean variance optimisation results (see Table \ref{tab:unconstrainedMV}) and solutions V and VIII of the constrained mean variance optimisation results (see Table \ref{tab:linearcons}) are geared versions of the optimal risky portfolio, it is appropriate that the \citet{goltsandjones2009} conditioned matrix was applied to this solution (See Figure \ref{fig:normal_distr_random_sims_AW}(a)). 

Unconstrained simulations without the use of covariance shrinkage show that only 3.8\% of optimal risky portfolios simulations had better (lower) $\alpha$-weight angles than the equally weighted portfolios, and were on average 27.5$^\circ$ worse than the equally weighted portfolio $\alpha$-weight angles. This demonstrate that the naive treatment results in equally-weighted portfolios, on average, out performing the mean-variance optimal portfolios. This confirms the distribution free perspective presented above, that the out performance of equally weighted benchmarks in simulations is due to selecting sub-optimal optimisation out-comes under estimation uncertainty. 

Second, when considering the $\alpha$-weight angle of constrained mean variance optimisation solutions VI and VII of Table \ref{tab:linearcons} compared to that of equally weighted portfolios, only 3.4\% and 3.2\% of optimal risky portfolios simulations had better (lower) $\alpha$-weight angles than the equally weighted portfolios for solutions VI and VII respectively. Additionally, the $\alpha$-weight angles were both 27.6$^\circ$  worse than the equally weighted portfolio's $\alpha$-weight angle respectively. Again, confirming the result that naive simulations would suggest that equally weighted portfolio's would outperform the mean variance portfolio's.  

Finally, simulations were then repeated to assess whether the covariance shrinkage methodology proposed by \citet{goltsandjones2009} for the optimal risky portfolio ($\theta_\alpha)$ allows for consistently smaller $\alpha$-weight angles than those produced by equally-weighted portfolios, in a normally distributed environment, and in the presence of constraints. When considering constrained mean variance solutions VI and VII of Table \ref{tab:linearcons}, which are not simply geared versions of the optimal risky portfolio $\theta_\alpha$ (Figure \ref{fig:normal_distr_random_sims_AW}(a)) because of the linear constraints, similar results can be seen; these are evident in Figure \ref{fig:normal_distr_random_sims_AW}(b) below. As soon as some conditioning is introduced, 100\% of simulations show better $\alpha$-weight angles than those produced by the equally weighted portfolios. This implies that 100\% of the simulations show improved covariance matrix conditioning, given the fact that the the matrix condition number provides an upper bound to the $\alpha$-weight angle (see equations \ref{eq:12} and \ref{eq:bahangle}).  

This confirms the theoretical implications, that only the correct formulation of conditioning creates appropriate optimal outcomes under linearly constrained quadratic optimisation. That the equally weighted portfolio, in the distribution free setting, are typically sub-optimal relative to the $\alpha$-angle shrunk covariances, and that this has been confirmed under simulation for the case of normal distributed returns. 

\subsection{Simulated Portfolio Results} \label{ssec:sims}

Various alternatives for shrinkage estimates are available, and this section intends to briefly introduce the benefits of the \citet{goltsandjones2009} methodology, through real data simulations.
We make use of monthly total return data from all stocks included in the FTSE/JSE All Share Index from January 1996 to August 2016. This includes 332 stocks, all with varying times of listing/de-listing. A set of 1000 simulations is performed. %, is run twice. One set assumes we know nothing about the expected return vector, and use historical means, the other set assumes we have perfect forecast ability of expected returns.
Each simulation follows the steps outlined in Table \ref{tab:simulation}.

\begin{table}[ht]% \sidecaption
\footnotesize
\centering
\begin{tabular}{p{0.46\textwidth}}
\toprule
{$\alpha$-Angle Simulation Methodology} \\
\midrule
Description \\
\midrule
{\bf1. Start date}: A random start date is selected and all stocks with a 10 year history following this date are noted. \\
\hline
{\bf 2. Investment universe}; 10 stocks are randomly selected from the full set of stocks in the randomly selected 10 year window. These form the data for this simulation (120 x 10 data frame) \\
\hline
{\bf 3. Simulation data window}: The first five years of the data is used in the simulation to find the sample covariance matrix. The shrunk covariance matrix of each shrinkage methodology is then computed at each shrinkage intensity ($k$ between 0 and 1, in increments of 0.01). %Please note for the first set of 1000 simulations,  
This five year history is also used to calculated the expected return vector, by merely using the historical mean. \\
\hline
{\bf 4. Calculate $\theta_{\alpha}$}:  The optimal risk portfolio weights are found for each of the shrinkage methodologies, for varying $k$. \\
\hline
{\bf 5. Calculate the $\alpha$-weight angle}: This is found for the varying shrinkage methodologies, to see how aligned the portfolio weights are to the forward returns. The forward returns is the performance of those assets in 2 months time. \\
\hline
{\bf 6. Out-of-sample Performance}: The out-of-sample portfolio performance of the Optimal Risky portfolio, for each of the shrinkage methods, at varying shrinkage intensity ($\kappa$) is calculated. This is performed by multiplying the optimiser weights (which uses the various covariance matrices as inputs), with the performance of each asset in 2 months time. From this performance, we find the out-of-sample Sharpe ratio (SR), and a probabilistic Sharpe ratio (PSR) (see \citet{deprado}). \\
\hline
{\bf 7. Step forward}: We then roll the five year period forward one month, until 5 years (minus 1 month) of out-of-sample performance is calculated, for each of the shrinkage methodologies, at varying $k$. \\
\bottomrule
\end{tabular}
\caption{Simulation steps for the demonstration of the out-of-sample behaviour of the different shrinkage methods and their impact on the $\alpha$-weight angle. In step 5, we delay the implementation by one month and are not using the asset's direct price 1 month later, or the return over the following months performance, because this would not be practically possible for managers to implement from a timing perspective. For practical relevance a month is skipped to allow for the real world mechanics of fund management implementation. It would be equivalent to introducing a look-a-head bias in the simulation. In addition, we note that in step 6, the probabilistic Sharpe ratio  is included because it provides us with a confidence level of a particular Sharpe ratio} \label{tab:simulation}
\end{table}

An important adjustment to note is that alpha-weight angles  can be greater than 90 degrees (between 90 and 180 degrees) in real world data. This infrequently occurs when there are large negative expected return vectors, or large negative correlations. The \citet{goltsandjones2009} methodology does not make provision for this and is addressed in the simulations by applying a cap on the alpha-weight angles. This means that any alpha-weight angle greater than 90 degrees, is set to the maximum value of 90 degrees.

Seven well known shrinkage methodologies are tested in this simulation. The range of shrinkage methods in the literature is significant, and this list is by no means exhaustive. For this reason we refer the interested reader to one of the many review papers given in the introduction. In terms of indicative methods we focus on those given by \citet{ledoito95} and \citet{ledoitwolf03,ledoitwolf04b}. This is because we are just trying to show from this sample, the competitive nature of our proposed methodology and when we believe it should be used. The seven different methodologies we consider are given in Table \ref{tab:shrinkage}.

\begin{table}[h!]% \sidecaption
\footnotesize
\centering
\begin{tabular}{lp{0.4\textwidth}}
\toprule
\multicolumn{2}{c}{Selected Shrinkage Methodologies Considered} \\
\midrule
Acr. & Description \\
\midrule
Diag. & {\bf Diagonal Matrix}: Shrinkage moves away from the sample covariance matrix toward the diagonal covariance matrix, $\diag{\Sigma}$, at intensity $k$: $\tilde \Sigma = (1-k)\Sigma + k \diag{\Sigma}$. The diagonal covariance matrix $\diag{\Sigma}$ is one where the diagonal of the sample covariance matrix is preserved, but the off-diagonal terms are all set to zero \citep{ledoito95} (Appendix B2). \\
CCM & {\bf Constant Correlation Matrix}: Shrinkage moves away from the sample covariance matrix toward the constant correlation covariance matrix, $\Sigma^{CCM}$, at intensity $k$: $\tilde \Sigma = (1-k)\Sigma + k\Sigma^{CCM}$. Here $\Sigma^{CCM}$ the diagonal of the sample covariance matrix is preserved, but the off-diagonal terms are all set to the average of all the stock covariances \citep{ledoitwolf04a}: $\Sigma_{ij}^{CCM} = \left\{ \Sigma_{ii}, i = j; \frac{1}{n(n-1)/2}\sum_{i=1}^n \Sigma_{ij}, i \neq j \right\}.$ \\
OPM & {\bf One Parameter Matrix}: Shrinkage moves away from the sample covariance matrix toward the one parameter covariance matrix, $\Sigma^{OPM}$, at intensity $k$: $\tilde \Sigma = (1-k)\Sigma + k \Sigma^{OPM}$. Here $\Sigma^{OPM}$ has the diagonal terms set to the the average variance, while the off-diagonal terms are all set to zero \citep{ledoitwolf04b}: $\Sigma_{ij}^{OPM} = \left\{ \frac{1}{n} \sum_{i=1}^n \Sigma_{ii}, i = j; 0, i \neq j \right\}$. \\
TPM & {\bf Two Parameter Matrix}: Shrinkage moves away from the sample covariance matrix toward the two parameter covariance matrix, $\Sigma^{TPM}$, at intensity $k$: $\tilde \Sigma = (1-k)\Sigma + k\Sigma^{TPM}$. Here $\Sigma^{TPM}$ is one where diagonal terms of the sample covariance matrix are all set to the the average variance, and the off-diagonal terms are all set to the average of all the stock covariances \citep{ledoito95} (Appendix B.1): $\Sigma_{ij}^{TPM} = \left\{ \frac{1}{n}\sum_{i=1}^n \Sigma_{ii}, i = j; \frac{1}{n(n-1)/2}\sum_{i=1}^n \Sigma_{ij};i \neq j \right\}$. \\
OFMM & {\bf One Factor Market Model}: Shrinkage moves away from the sample covariance matrix toward the one factor market model covariance matrix, $\Sigma^{OFMM}$, at intensity $k$: $\tilde \Sigma = (1-k)\Sigma + k(\Sigma^{OFMM})$. Here $\Sigma^{OFMM}$ is one where the diagonal is preserved (due to the idiosyncratic volatility of the residuals), but the off-diagonal terms are shrunk to the cross-sectional average of all the random variables: $\Sigma_{ij}^{OPMM}= (s^2_m \beta \beta' + \epsilon)_{ij}$ \citep{ledoitwolf03}. Here $\beta$ is the $n \times 1$ matrix of market factor loadings, $\epsilon$ a $n \times n$ diagonal matrix of residual variance estimates, and $s^2_m$ the market sample variance. \\
AOCS & {\bf Asymptotically Optimal Convex Shrinkage}: The covariance matrix is a linear shrinkage between the sample covariance matrix and the identity matrix: $\tilde \Sigma = (1-\alpha)\Sigma +\alpha(\tau\Sigma)$ \citep{ledoitwolf04b}.Here $\tau$ is the average sample variance and $\alpha \in [0,1]$ is the intensity parameter computed using $\alpha = \frac{1}{\trace([\Sigma-\tau]^2)} \frac{1}{N}\sum_{i=1}^{N} \trace([z_iz_i^T - \Sigma]^2)$. This is the only method investigated in this paper, where the shrinkage intensity is set (and $k$ not varied) in the simulations. \\
G\&J &  {\bf Golts \& Jones $\alpha$-angle minimization}: The covariance matrix is a linear shrinkage between the sample covariance matrix and the identity matrix, with the $\alpha$-weight angle informing the degree of shrinkage as follows (see Equation \eqref{eq:shrinkcov}), repeated here: $\tilde \Sigma = \frac{k}{\cos \varphi} I + \left({1 -  \frac{k}{\cos \varphi}}\right) \Sigma$. Here $k$ is some quantity $\cos (\varphi_k)$ for a targeted optimal angle $\varphi_k \in[\varphi,\pi/2]$ between the risky portfolio and the expected returns.\\
\midrule
\end{tabular}
\caption{Shrinkage method descriptions for the methods compared in the comparative portfolio simulations (Figures \ref{fig:sln5and8}, \ref{fig:sln6} and \ref{fig:sln7}).} \label{tab:shrinkage}
\end{table}

\begin{figure*}[ht!]
    \centering
    \subfloat[][Sharpe Ratio]{
    \includegraphics[width=0.45\textwidth]{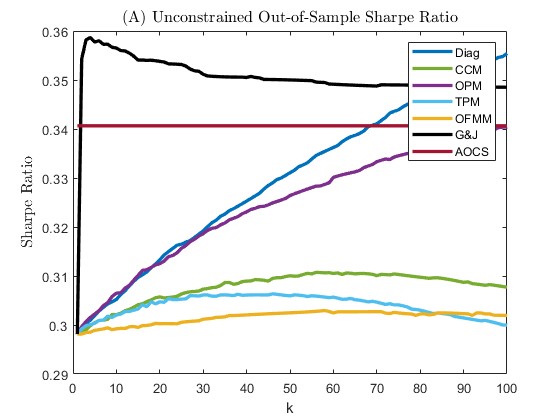} \label{fig:sln5and8_A}
    } 
    \subfloat[][Probabilistic Sharpe Ratio]{\includegraphics[width=0.45\textwidth]{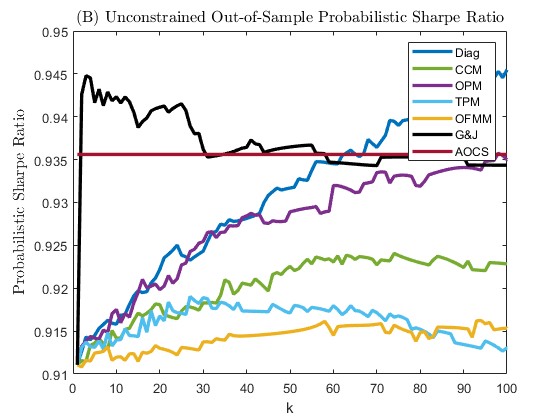} \label{fig:sln5and8_B}
    } \\
     \subfloat[][Average $\alpha$-weight angle]{\includegraphics[width=0.45\textwidth]{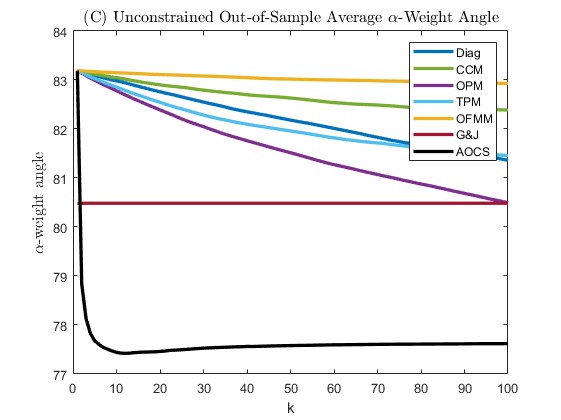}  \label{fig:sln5and8_C}
    }
     \subfloat[][Condition Number improvement]{\includegraphics[width=0.45\textwidth]{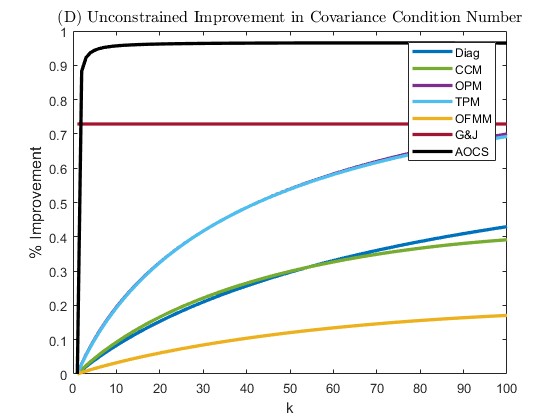} \label{fig:sln5and8_D}
    } 
    \caption{The historic simulations results for the unconstrained solutions (Table \ref{tab:unconstrainedMV}). Fig. \ref{fig:sln5and8_A} shows the average out-of-sample Sharpe ratios, Fig. \ref{fig:sln5and8_B} shows the average out-of-sample probabilistic Sharpe ratio's, Fig. \ref{fig:sln5and8_C} shows the average $\alpha$-weight angle, and Fig. \ref{fig:sln5and8_D} shows the average percentage improvement of the condition number of the covariance matrix; all across 1000 simulations, for varying $k$, for each shrinkage methodology given in Table \ref{tab:shrinkage}. The G\&J appears to deliver returns most aligned to future asset returns, generates higher Sharpe ratios, at high confidence levels, and improves the condition number of the covariance matrices significantly.}
    \label{fig:sln5and8}
\end{figure*}

\begin{figure*}[h!]
    \centering
       \subfloat[][Sharpe Ratio]{
    \includegraphics[width=0.45\textwidth]{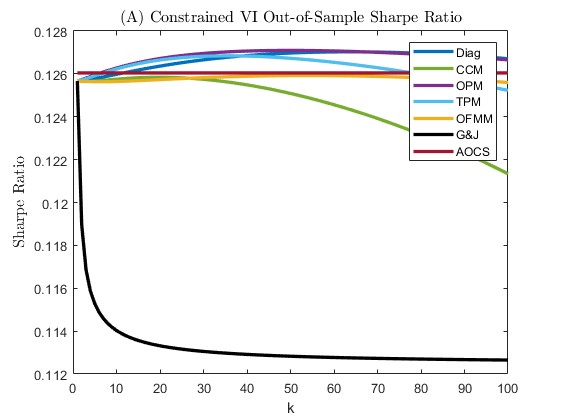} \label{fig:sln6_A}
    } 
    \subfloat[][Probabilistic Sharpe Ratio]{\includegraphics[width=0.45\textwidth]{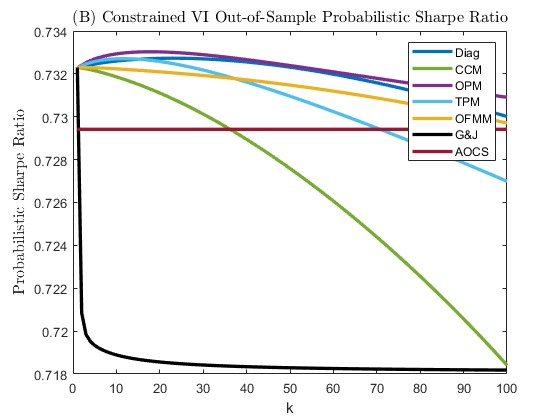} \label{fig:sln6_B}
    } \\
     \subfloat[][Average $\alpha$-weight angle]{\includegraphics[width=0.45\textwidth]{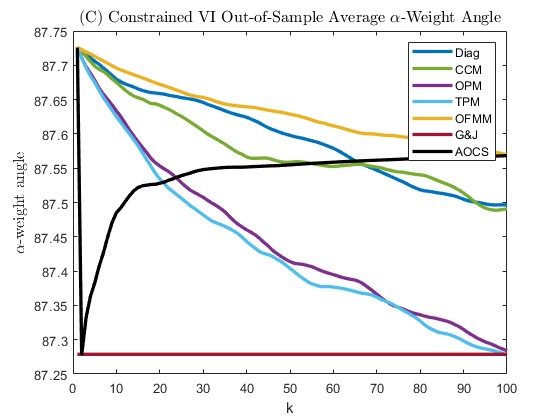}  \label{fig:sln6_C}
    }
     \subfloat[][Condition Number improvement]{\includegraphics[width=0.45\textwidth]{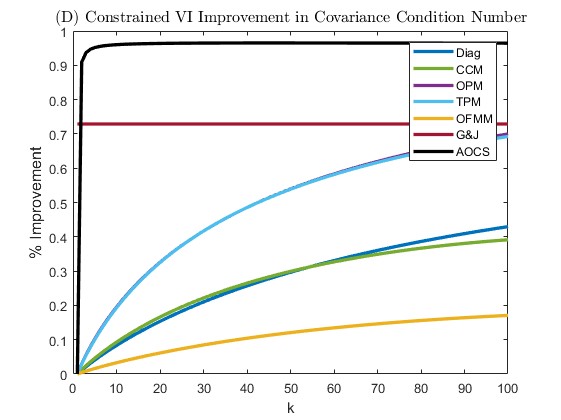} \label{fig:sln6_D}
    } 
    \caption{The historic simulation results for the constrained solution VI (Table \ref{tab:linearcons}) are shown. Fig.\ref{fig:sln6_A} shows the average out-of-sample Sharpe ratios, Fig. \ref{fig:sln6_B} shows the average out-of-sample probabilistic Sharpe ratios, Fig. \ref{fig:sln6_C} shows the average $\alpha$-weight angle, and Fig. \ref{fig:sln6_D} shows the average percentage improvement of the condition number of the covariance matrix; again across 1000 simulations, for varying $k$, for each shrinkage methodology given in Table \ref{tab:shrinkage}. While G\&J methodology improves the condition number of the covariance matrices, and see's strong $\alpha$-weight angles at small $k$, the methodology does not appear superior under this solution.}
    \label{fig:sln6}
\end{figure*}

\begin{figure*}[h!]
    \centering
    \subfloat[][Sharpe Ratio]{
    \includegraphics[width=0.45\textwidth]{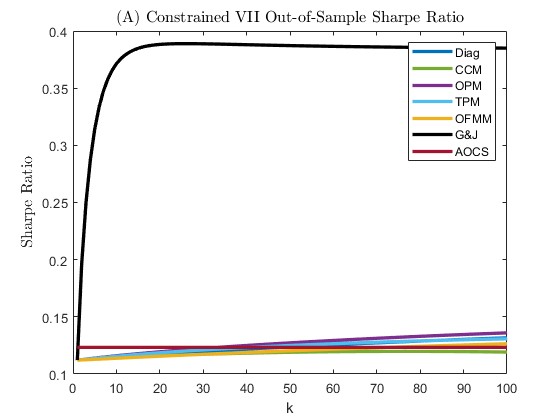} \label{fig:sln7_A}
    } 
    \subfloat[][Probabilistic Sharpe Ratio]{\includegraphics[width=0.45\textwidth]{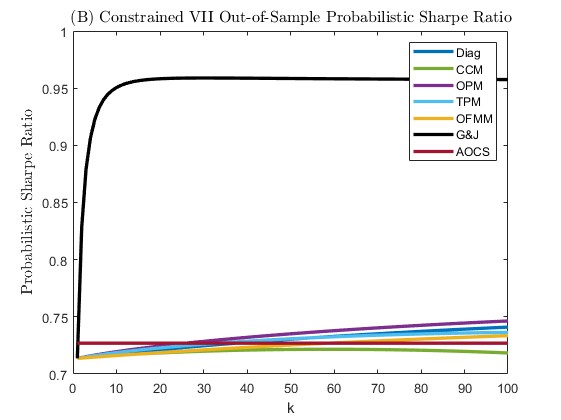} \label{fig:sln7_B}
    } \\
     \subfloat[][Average $\alpha$-weight angle]{\includegraphics[width=0.45\textwidth]{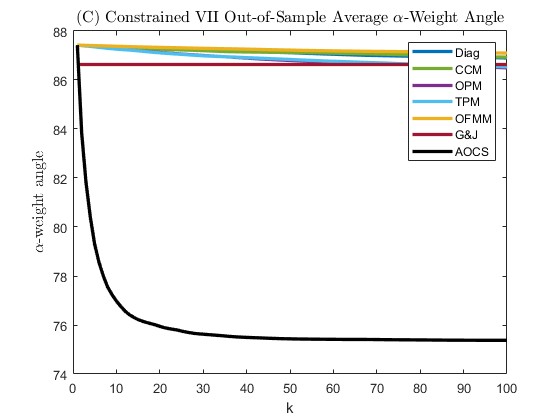}  \label{fig:sln7_C}
    }
     \subfloat[][Condition Number improvement]{\includegraphics[width=0.45\textwidth]{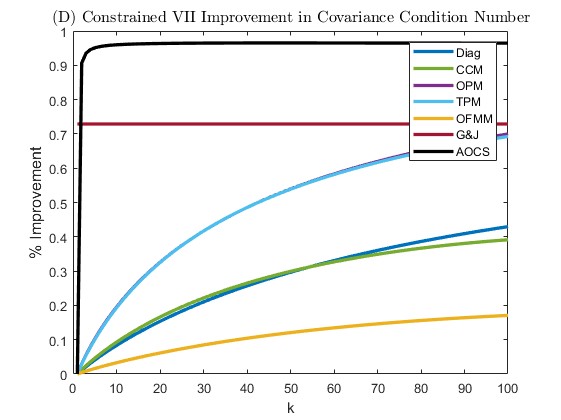} \label{fig:sln7_D}
    } 
    \caption{The historic simulation results for the constrained solution VII  (Table \ref{tab:linearcons}) are shown. Fig . \ref{fig:sln7_A} shows the average out-of-sample Sharpe ratios, Fig. \ref{fig:sln7_B} the average out-of-sample probabilistic Sharpe ratios across 1000 simulations, for varying $k$, for each shrinkage methodology, Fig. \ref{fig:sln7_C} the average $\alpha$-weight angle, and Fig. \ref{fig:sln7_D} shows the average percentage improvement of the condition number of the covariance matrix across 1000 simulations, for varying $k$, for each shrinkage methodology given in Table \ref{tab:shrinkage}. G\&J appears to deliver returns most aligned to future asset returns,generates higher Sharpe ratios, at high confidence levels, and improves the condition number of the covariance matrices significantly.} \label{fig:sln7}
\end{figure*}

Figure \ref{fig:sln5and8} provides the simulation results for the unconstrained solutions (Table \ref{tab:unconstrainedMV}). This includes the constrained solutions (V and VIII from Table \ref{tab:linearcons}) which are simply geared versions of each other. The results are shown across 1000 simulations, for varying $k$, for each shrinkage methodology. This figure shows the average out-of-sample Sharpe ratios (Figure \ref{fig:sln5and8_A}), the average out-of-sample probabilistic Sharpe ratio's (Figure \ref{fig:sln5and8_B}), the average $\alpha$-weight angle (Figure \ref{fig:sln5and8_C}) and the average percentage improvement of the condition number of the covariance matrix (Figure \ref{fig:sln5and8_D}). Here the $\alpha$-weight angle is the angle between the "shrunk" weights and the forward looking returns, showing which methodology will deliver performance most in line with future returns. G\&J appears to deliver returns most aligned to future asset returns. We see that the G\&J methodology generates higher Sharpe ratios, at high confidence levels, especially around smaller $k$ estimates, and the G\&J methodology improves the condition number of the covariance matrices significantly.

Figure \ref{fig:sln6} shows the simulations results for the constrained solution VI from Table \ref{tab:linearcons}. The figure shows the average out-of-sample Sharpe ratios (Figure \ref{fig:sln6_A}), the average out-of-sample probabilistic Sharpe ratios (Figure \ref{fig:sln6_B}), 
the average $\alpha$-weight angle (Figure\ref{fig:sln6_C}) and the average percentage improvement of the condition number of the covariance matrix (Figure \ref{fig:sln6_D}) across 1000 simulations, for varying $k$, for each shrinkage methodology. We see that the G\&J methodology does not tend to generate higher sharpe ratios using solution VI. Again, we note that the $\alpha$-weight angle is the angle between the``shrunk" weights and the forward looking returns, showing which methodology will deliver performance most in line with future returns. While the $\alpha$-weight angle does decrease very slightly, the G\&J methodology does not appear to be most favourable in this regard. 

Figure \ref{fig:sln7} demonstrates the simulation results for the constrained solution VII from Table  \ref{tab:linearcons}. Here the average out-of-sample sharpe ratios are in Figure \ref{fig:sln7_A}, the average out-of-sample probabilistic Sharpe ratios in Figure \ref{fig:sln7_B}, the average $\alpha$-weight angle in Figure \ref{fig:sln7_C}, and the average percentage improvement of the condition number of the covariance matrix is given in Figure \ref{fig:sln7_D}.
The historic simulations are across 1000 simulations following the methodology outlined in Table \ref{tab:simulation}, for varying $k$, and for each shrinkage methodology from Table \ref{tab:shrinkage}. We see that the G\&J methodology generates higher Sharpe ratios, at high confidence levels, especially around smaller $k$ estimates. Again, we can see which methodology will deliver performance most in line with future returns because the $\alpha$-weight angle is the angle between the ``shrunk" weights and the forward looking returns. The G\&J method appears to deliver returns most aligned to future asset returns, and improves the condition number of the covariance matrices significantly.

In summary, what one can conclude from the simulations shown in figures \ref{fig:sln5and8}, \ref{fig:sln6} and \ref{fig:sln7}, is that under unconstrained solutions and constrained solutions V, VII and VIII the \citet{goltsandjones2009} method appears superior in delivering portfolios that are better aligned to the returns of the assets, and generating out-of-sample performance with superior Sharpe ratios. 

However, under constrained Solution VI, this does not appear to be the case. Solution VI implements a minimum return threshold, and solutions naturally tend to solve for weights that generate that minimum return exactly. This means that all solutions are solving for the same return, and improved Sharpe ratio's are only being generated from the standard deviation, which the G\&J methodology does not necessarily solve for. Moving the portfolio weights closer to the expected return vector (shrinking the $\alpha$-weight angle) is not beneficial when the return benefit is not there.  

Importantly, we tested this methodology naively using historical return means as the expected return vectors, and even with this large degree of naivety around the future asset returns, this methodology returns weights more closely resembling future asset returns. So while using historical returns is a reasonable enough forward estimate for future returns to make this method beneficial in practice, when managers have more insight, even intermittently, into future expected returns, it is clear that this methodology will be even more useful.

%In choosing the shrinkage intensity ($k$) of this methodology, it seems clear than a $k$ of around x appears to generate the best deflated sharpe ratio's most consistently, at high confidence levels (significant p-values).  
%In the case where one has no inclination about future asset returns, this methodology is however not superior and the diagonal matrix shrinkage methodology appears superior, especially at high shrinkage intensity levels, when the covariance matrix is far from the sample covariance matrix. We therefore conclude what is intuitively clear from the mathematics, that if you have an inclination of where expected returns will be in the future, it is worthwhile rotating the weight vector toward the expected return vector for superior out-of-sample risk adjusted portfolio's.  

\section{Conclusions}

% \label{eq:shrinkcov} -- proposed shrinkage
%  \label{eq:37} -- constrained alpha weight angle
% \label{eq:minriskgear5} -- key optimisation
%  \label{eq:varphi01}   -- equivalence of the optimisation 
Mean-variance portfolio decisions that combine prediction and optimisation have been shown to have poor empirical performance for a variety of shrinkage predictors under different distributional assumptions that combined mean-variance optimisation with reasonable departures from Gaussianity \citep{Paskaramoorthyetal2021}. Here we directly consider both constrained and unconstrained mean-variance portfolio construction and performance in a general setting that does not have strong distributional assumptions and we then confirm these theoretical results using a simple Gaussian simulation. This is made possible by casting mean-variance portfolio theory into a geometric setting.

To present the geometric perspective we consider the upper bound of the $\ve \alpha$-weight angle as introduced by \citet{goltsandjones2009}. First, we show how this result arises as a result of the Kantorovich inequality in Theorem \ref{thm:kant}. Despite being framed in the mean-variance setting the $\ve \alpha$-angle approach to covariance shrinkage makes no distributional assumptions about the data generating processes. \citet{goltsandjones2009} only considered unconstrained optimisation problems when demonstrating the relationship between the $\ve \alpha$-weight angle and the condition of the covariance matrix; as in optimisations I, II, III and IV in Table \ref{tab:unconstrainedMV}. Then, we show that this result can be weakened and generalised when gearing constraints are added; as in the optimisations V, VI, VII and VIII in Table \ref{tab:linearcons} using the extension of Bauer and Householder given by Theorem \ref{thm:bah}. 

When looking at the results of various forms of mean variance optimisation problems, we note that results of the unconstrained optimisations: I,II, III and IV, can simply be thought of as portfolios combined with a risk-less portfolio and hence generate Pareto surfaces equivalent to the Capital Market Line with a risk fee rate set to 0\%. Interestingly, these optimisations (I, II, III and IV) all have implicit gearing, as imposed by the choice of $\alpha_0$, $\sigma_0$ and $\gamma$, despite that there is no outright, or explicit gearing constraint. 

We see that optimisation V is equivalent to I because the choice of gearing, $g_0$, is equivalent to a choice of portfolio risk, $\sigma_0$. Optimisations I,II,III,IV as well as gearing constrained optimisations V and VIII, all have the same investment directions; in the direction of the optimal risky portfolio (i.e. they are all geared versions of the optimal risky portfolio). The optimal risky portfolio is always fully invested, and gearing can subsequently be applied, which will make the portfolio move up the Capital Market Line and be the tangent portfolio for the efficient frontier where the level of gearing is a constraint. 

It is the fact that these solutions are in the direction of the optimal risk portfolio that makes the Kantorovich inequality (Theorem 1) appropriate for proving the upper bound of the $\ve \alpha$-weight angle (using the lower bound of the cosine of this angle). From a geometric perspective there is no tension between the gearing and the $\ve \alpha$ direction in these cases, and there is only a tension between the poor conditioning of the covariance matrix and the direction of $\ve \alpha$. This insight lead to the idea of directly minimising the $\ve \alpha$-angle by shrinking the covariance because for these optimisations this direction is generated by the covariance matrix: {\it i.e.} by the vector $\Sigma^{-1} \ve \alpha$. This is not equivalent to including a return maximisation constraint.

However, when adding an explicit gearing restriction to certain problems, such as solutions VI and VII, gearing can actually mimic poor conditioning and the portfolios will have investment directions different from those optimisations without a gearing constraint. From the two fund separation theorems the resulting optimal portfolio is a weighted average of the global minimum variance portfolio and the optimal risky portfolio. We see that naively applying gearing to any optimisation other than optimisation IV (the Sharpe optimal portfolio that generates the optimal risky portfolio) and VIII (the return maximisation) will drag the portfolios out of alignment with $\ve \alpha$ towards the global minimum variance portfolio. 

Using fund separation theorems the global minimum variance portfolio can be factored from the optimal portfolio Eqn.(\ref{eq:minriskgear5}). Geometrically there is now a tension between the gearing, the $\ve \alpha$ direction, and the impact of poor conditioning. However, the global minimum variance portfolio is shown to be purely a function of the covariance matrix, and is a fully invested portfolio that sums to one; it's gearing can be managed separately from the minimisation of the $\ve \alpha$-angle. This key insight allows us to retain the idea of minimising the $\ve \alpha$-angle.

The gearing constrained optimal portfolio has a weakened upper bound that is now given by Theorem \ref{thm:bah} \citep{bauerhouseholder1960}, and not Theorem \ref{thm:kant} (the Kantorovich inequality) because of the impact of the constraint. With improved conditioning of the covariance matrix we see we are still able to reduce the upper bound on the angle between the optimal risky portfolio and $\ve \alpha$, which would in turn reduce the overall angle between the optimal portfolio and the direction of $\ve \alpha$ and improve overall portfolio performance out-of-sample. In fact, we note for the constrained optimisation from Eqn.(\ref{eq:minriskgear5}) that a minimisation over the total $\alpha$-angle $\varphi$ is equivalent to minimising the angle between the optimal risky portfolio and the direction of expected returns. This extends the unconstrained result and motivates why the combination of an $\ve \alpha$-angle minimisation is equally useful both in the unconstrained and constrained cases. To minimise the overall angle between the optimal portfolio and the expected returns we can minimise the angle between the optimal risky portfolio and the expected returns, Eqn.(\ref{eq:riskyretmin1}). 

What this means practically is that even when faced with poor forecast ability in the presence of estimation uncertainty, uncertainty that is amplified by the inversion of the covariance matrix, the choice is not a binary one between {\it e.g.} an equally weighted, or risk-weighted portfolio, and the mean-variance solution; but rather there is a smooth deformation of the solution between the traditional mean-variance and the risk-weighted or equally-weight solutions, that can enter the problem via constraints where optimality is dependent on the conditioning of the covariance and the constraints. This suggest that there is a regularisation parameter that can be optimally chosen for a given use case and dataset. 

Towards this end we again look at the spherical uncertainty region model for $\ve \alpha$, and used it to motivate a simple robust optimisation framework. This motivates the use of covariance matrix shrinkage that explicitly reduces the angle between optimal risky Sharpe ratio portfolio, the optimal risky portfolio, and the direction of expected returns, and is given in Eqn.(\ref{eq:shrinkcov}). We shrink the covariance matrix using the $\ve \alpha$-angle minimisation with some regularisation parameter $k$ which is thought of as some optimal regularisation angle $\varphi_k$. 

By regularising the covariance matrix to constrain the direction of the optimal risk portfolio we achieve the overall $\ve \alpha$-angle minimisation regularisation as in Eqn.(\ref{eq:shrinkcov}) that is consistent with reducing the impact of the estimation uncertainty, as expressed by the upper bound in Eqn.(\ref{eq:37}) from Theorem \ref{thm:bah}. This choice demonstrates the emergence of special cases, such as the equally weight portfolio solutions as the limiting case of the global minimum variance portfolio, and how this is optimally mixed within the linearly constrained mean-variance solution.

This is demonstrated for the case of normal returns in Section \ref{ssec:alphadepend}. Where we show that in the case of niave unconstrained portfolio optimisation that the equally-weighted portfolio's performance better than the mean-variance optimal portfolios. However, when applying the method first proposed by \cite{goltsandjones2009} this is no longer the case, the mean-variance portfolio's would on average out-perform the equally weighted portfolio's. Similarly in the presence of linear constraints, when the shrinkage hyper-parameters are appropriate selected one can ensure that the constrained mean-variance portfolio's, on average, out-perform the equally weighted portfolio's. This is consistent with the distribution free general result used to motivate the method of \cite{goltsandjones2009}. 

This is in turn confirmed using historic simulation in Section \ref{ssec:sims}. Here we find that under unconstrained solutions and constrained solutions V, VII and VIII that the method of \citet{goltsandjones2009} delivers portfolios that are better aligned to the returns of the assets, and generating out-of-sample performance with superior Sharpe ratios than the other methods considered.  However, we note that under constrained solution VI, this is not the case because that particular solution naturally tends to solve for weights that generate that minimum returns exactly, and as such moving the portfolio weights closer to the expected return vector is not beneficial despite improve the conditioning of the covariance matrix.

Practically, when a fund manager is extremely uncertain about the covariance between the assets, and uncertain about the returns, but one has a gearing constraint, then one should be holding Eqn.(\ref{eq:uncertain1}) where the broad direction of expected returns is still of value. However, if one has a sense for the relative variances of assets, then instead of the diagonal matrix $I$ in Eqn.(\ref{eq:shrinkcov}), one could use the diagonal covariance matrix with zero off-diagonals. The key realisation is not that one is optimally shrinking the covariance matrix to the diagonal matrix to dilute correlations to ameliorate the impact of estimation uncertainty under covariance matrix inversion, but rather that one is optimally distorting the covariance matrix so as to optimally exploit the geometry of the control problem itself in the presence of constraints. 

\section*{Ackowledgements}
We thank Dave Bradfield for thoughtful and inspiring discussions. 

\section*{Conflict of Interest}
The authors declare there is no conflict of interest.

\bibliographystyle{elsarticle-harv}
\setcitestyle{authoryear,open={(},close={)}} %Citation-related commands
\bibliography{LDTG-References}

\appendix

\section{Kantorovich Inequality} \label{app:Thrm1}

Following \cite{kant21964} and \cite{kant11985}. Let $A \in M_n$ be Hermitian and positive definite; let $\rho_1$ and $\rho_n$ be its smallest and largest eigenvalues. Let there be a non-zero vector $\vec x \in \C^n$ with a Euclidean norm. Consider two positive semi-definite matrices $\rho_nI - A$ and $A-\rho_1I$, and a positive definite matrix $A^{-1}$. These are Hermitian matrices and commute; their product is Hermitian and positive semi-definite. Therefore, for any non-zero complex valued vector $\vec x$ with a complex conjugate transpose $\vec x^{_H}$
\begin{align}\label{eq:thrm1-2}
 0 &\leq \vec x^{_H}(\rho_nI-A)(A-\rho_1I)A^{-1} \vec x \nonumber \\ &= \vec x^{_H}\left({(\rho_1 + \rho_n)I - \rho_1\rho_nA^{-1}-A}\right) \vec x
\end{align}
and so $\vec x^{_H}A \vec x + \rho_1\rho_n(\vec x^{_H}A^{-1} \vec x) \leq (\rho_1 + \rho_n)(\vec x^{_H}\vec x)$. If $t_0 = \rho_1\rho_n(\vec x^{_H}A^{-1} \vec x)$ then
$t_0(\vec x^{_H}A \vec x) \leq t_0(\rho_1 + \rho_n)(\vec x^{_H}\vec x) - t_0^2$. The function $f(t) = t(\rho_1+\rho_n)(\vec x^{_H}\vec x)- t^2$ is concave and has a critical point at $t=\frac{1}{2}(\vec x^{_H} \vec x)(\rho_1+\rho_n)$ where it has a global maximum. Therefore $f(t_0) \leq \frac{1}{4}(\vec x^{_H} \vec x)^2(\rho_1+\rho_n)^2$ and it then follows that $\rho_1\rho_n(\vec x^{_H}A^{-1} \vec x)(\vec x^{_H}A \vec x) \leq \frac{1}{4}(\rho_1+\rho_n)^2(\vec x^{_H}\vec x)^2$. This is the Kantorovich inequality:
\begin{equation}\label{eq:thrm1-1}
     (\vec x^{_H}A \vec x)(\vec x^{_H}A^{-1}\vec x) \leq \frac{(\rho_1 + \rho_n)^2}{4\rho_1\rho_n} (\vec x^{_H} \vec x)^2 .
\end{equation}
The inequality now holds by construction for all $\vec v \in \R^n \backslash\{0\}$ as in Theorem \ref{thm:kant} where the Hermitian transpose $H$ becomes a matrix transpose $T$.

%R.A. Horn, C.R. Johnson, Matrix Analysis, Cambridge University Press, London, 1985.%

\section{Bauer and Householder Inequality} \label{app:Thrm2}

A very elegant direct proof is given in \cite{huangandzhou}. Here we rather follow \cite{bauerhouseholder1960} to motivate the result. Let $A$ be a non-singular matrix, and $\vec x$ any vector with Euclidean norm. When A has a Euclidean bound norm\footnote{For singular values $\sigma^2_i(A) = \lambda_i(A^{_H}A)$ with eigenvalues $\lambda_i$ the Euclidean norm is a scalar $\|A \|=\max_{\|x\|=1} \|A \vec x  \|= \sigma_{_\mathrm{max}}(A)$.} and if the inverse of A exists then there is an associated condition number\footnote{The condition number or Euclidean condition is a scalar $\kappa(A) = \|A  \|\cdot \|A^{-1} \|= \frac{\sigma_{\mathrm{max}}(A)}{\sigma_{\mathrm{min}}(A)}$} (also known as the Euclidean condition). Given any two unit vectors $\ve \xi$ and $\ve \eta$ then\footnote{For (Hermitian) complex conjugate transpose $H$.} $\ve \xi^{_H} \ve \eta = e^{i \theta} \cos \psi$ where the angle $\psi$ between the two unit vectors is restricted to $\psi \in [0,\pi/2]$. Consider matrices $\widetilde{A} = (A \ve \xi, A \ve \eta)$ and $\widetilde{M} = \widetilde{A}^{_H} \widetilde{A}$. If $\ve \zeta$ is another unit vector in the place of $\ve \xi$ and $\ve \eta$ and orthogonal to $\ve \xi$, then for some $\overline{M} = (\ve \xi, \ve \zeta)^{_H} A^{_H} A(\ve \xi ,\ve \zeta)$ we can define a matrix $Q = [ 1 ~e^{i\theta} \cos\psi; 0 ~e^{i\theta} \sin\psi]$ such that $\widetilde{M} = Q^{_H} \overline{M} Q$. From the Euclidean condition $\kappa(\widetilde{M}) \leq \kappa^2(Q) \kappa(\overline{M})$. If $\kappa(\overline{M}) \leq \kappa(A^{_H} A)$ then 
\begin{equation}\label{eq:thrm2-8}
\kappa^2(Q) = \kappa(Q^{_H}Q) = \frac{1 + \cos\psi}{1 - \cos\psi} = \cot^2\left( {\frac{\psi}{2}} \right).
\end{equation}
This means that for unit vectors $\ve \xi$ and $\ve \eta$ and $\widetilde{A} = (A \ve \xi, A \ve \eta)$ where $|\ve \xi^{_H} \ve \eta | \leq \cos(\psi)$ then $(\kappa(\widetilde{A}^{_H}A))^{\frac{1}{2}} \leq \kappa(A) \cot\sfrac{\psi}{2}$. Similarly, with $\widetilde{M} =(\ve \xi, \ve \eta)^{_H}M(\ve \xi, \ve \eta)$ and some positive definite $M$ then $\kappa(\widetilde{M}) \leq \kappa(M)\cot^2 \sfrac{\psi}{2}$. From the cosine angle between $A\ve \xi$ and $A\ve \eta$ and if $\widetilde{M} = [\widetilde{m}_{11} ~ \widetilde{m}_{12}; \widetilde{m}_{21} ~ \widetilde{m}_{22}]$ then:
\begin{equation}\label{eq:thrm2-12}
\left|\frac{\ve (A \ve \xi)^{_H} A \ve \eta}{\left \|A \ve \xi \right \|\cdot \|A \ve \eta \|}\right|^2  = \frac{\widetilde{m}_{12}\widetilde{m}_{21}}{\widetilde{m}_{11}\widetilde{m}_{22}} = 1-\frac{\det\widetilde{M}}{\widetilde{m}_{11}\widetilde{m}_{22}}.
\end{equation}
If $\widetilde{M}$ has characteristic roots $\tau$ and $\tau\kappa(\widetilde{M})$ then $\det \widetilde{M} = \tau^2 \kappa (\widetilde{M})$ and $(\widetilde{m}_{11}\widetilde{m}_{22})^\frac{1}{2} \leq \frac{1}{2}(\widetilde{m}_{11} +\widetilde{m}_{22}) = \frac{\tau}{2}(1+ \kappa(\widetilde{M}))$ so that \ref{eq:thrm2-12} is bounded above
\begin{align}\label{eq:thrm2-15}
\left|\frac{(A \ve \xi)^{_H} A \ve \eta}{\|A \ve \xi \|\cdot \|A \ve \eta \|}\right|^2 &\leq 1-\frac{4\kappa(M)}{(1+\kappa(M))^2} = 
\left( \frac{\kappa(M)-1}{\kappa(M) +1} \right)^2 \nonumber \\
&\leq \left( \frac{\kappa^2(A) \cot^2 \sfrac{\psi}{2} -1}{\kappa^2(A) \cot^2 \sfrac{\psi}{2} + 1} \right)^2. \nonumber
\end{align}
For $\kappa \geq 1$ the term $(\kappa -1)/(\kappa+1)$ is a monotonic function of $\kappa$ so we can choose some 
$\cot(\sfrac{\phi}{2}) = \kappa(A) \cot\sfrac{\psi}{2}$. The last term above, the squared quotient on the right, then becomes $\cos^2\psi$ with
$\ve \xi = \vec x \left \|\vec x \right \|^{-1} $ and $\ve \eta =  \vec y\left \| \vec y \right \|^{-1}$. Thus, we have shown that for any two non-null vectors $\vec x$ and $\vec y$, and any non-singular matrix A, that $|\vec y^{_H} \vec x| \leq \|\vec x\| \|\vec y\| \cos\psi$ for $\psi \in [0,\pi/2]$. This implies that
\begin{equation}\label{eq:thrm2-19}
    |(A \vec y)^{_H}A \vec x| \leq \| A \vec x \| \|A \vec y\| \cos\phi,
\end{equation}
for $\cot \sfrac{\phi}{2} = \kappa(A)\cot\sfrac{\psi}{2}$. 
Similarly, for any positive definite matrix $M$ this can then be extended to imply:
\begin{equation}\label{eq:thrm2-21}
|\vec y^{_H} M \vec x|^2 \leq (\vec x^{_H} M \vec x) (\vec y^{_H} M \vec y) \cos^2(\phi), 
\end{equation}
for $\cot^2\sfrac{\phi}{2} = \kappa(M)\cot^2\sfrac{\psi}{2}$. Now, consider an arbitrary rectangular matrix of linearly independent columns $Y$ so that $AY$ can be defined. 

The matrix $H = I-A^{_H}Y(Y^{_H}AA^{_H}Y)^{-1}Y^{_H}A$
is then a Hermitian idempotent annihilating\footnote{Solving systems of linear equations can be done using a sequence of Hermitian operators of the form $H$, each operating upon the previous residual $s$, to form a new residual $Hs$; this is the method of iterative projection \citep{householderbauer1960}.} $A^{_H}Y$. Using that we can choose some $M= A A^{_H}$, we change notation to $\vec r=A \vec s$ and $\vec x= AH \vec s = [I-MY(Y^{_H}MY)^{-1}Y^{_H}]\vec r$. If $\vec r$ is in the subspace of $MY$, then either $H \vec s=0$ or $\vec x$ is a null-vector, and then $\vec x^{_H}M^{-1}\vec x = \vec r^{_H}M^{-1}\vec x$. Using this, and that $H^2 = H$:
\begin{equation}\label{eq:thrm2-26}
\frac{\|H \vec s\|^2}{\| \vec s\|^2} =  \frac{(\vec r^{_H} M^{-1} \vec x)}{(\vec r^{_H} M^{-1} \vec r)}
 = \frac{|\vec r^{_H}M^{-1}\vec x|^2}{(\vec r^{_H}M^{-1} \vec r)(\vec x^{_H}M^{-1} \vec x)}.
\end{equation}
Since $\kappa(M^{-1}) = \kappa(M)$ and $\kappa(A^{_H}) = \kappa(A)$ then $
H=I-A^{_H}Y(Y^{_H}AA^{_H}Y)^{-1}Y^{_H}A$ and $\vec s=A^{-1} \vec r$. If $Y$ has only a single column $\vec y$ then $H=I-(A^{_H}\vec y)(\vec y^{_H}A)/||A^{_H}\vec y||$ and with some linear algebra it follows that
\begin{equation}\label{eq:thrm2-37}
\frac{\|H \vec s\|^2}{\| \vec s\|^2} = 1- \frac{|\vec r^{_H} \vec y|^2}{\|\vec y^{_H}A\| \|A^{-1} \vec r\|}.
\end{equation}

Using that $\vec x$ is orthogonal to $Y$ where the cosine of the angle between $\vec r$ and the subspace of $Y$ is $\cos(Y,\vec r) = {\|Y(Y^{_H}Y)^{-1}Y \vec r \|}/{\| \vec r \|}$ we have that $\cos(\vec x,\vec r) \leq [1-\cos^2(Y,\vec r)]^\frac{1}{2}$. Then we have that $\|Y(Y^{_H}Y)^{-1}Y^{_H}\vec r\| \geq \|\vec r\| \sin\psi$ which implies that $\|H \vec s\| \leq \|\vec s\| \cos\psi$. From this an optimal bound for \ref{eq:thrm2-26}, and thus \ref{eq:thrm2-37}, can be obtained when the space of $Y$ contains $\vec r$:
\begin{equation}\label{eq:thrm2-33}
\frac{||H \vec s||}{||\vec s||} \leq \frac{[\kappa^2(A)-1]}{[\kappa^2(A)+1]}.
\end{equation}
The bound $|\vec y^{_H}\vec r| \geq \|\vec y\| \|\vec r\| \sin\psi$ for implies $\|H\vec s\| \leq \|\vec s\| \cos\phi$. Now, if the equality $|\vec y^{_H} \vec r| = \|\vec y\| \|\vec r\| \sin\psi_0$ holds for $\psi_0 \in [0,\pi/2]$ then:
\begin{equation}
\frac{\| \vec y^{_H}A\| \|A^{-1} \vec r\|}{\|\vec y^{_H}\| \|\vec r\|} \leq \frac{\sin\psi_0}{\sin\phi} = \frac{1}{2}\left[{(\kappa + \kappa^{-1})+(\kappa - \kappa^{-1})\cos\psi_0}\right]. \nonumber
\end{equation}
Here the bound on the right decreases as $\psi_0$ increases because $\cos\psi_0$ decreases on $[0,\pi/2]$. It then follows that for any non-null vectors $\ve u$ and $\vec v$, and any non-singular matrix $A$, that $|\vec v^{_H} \vec u| \geq \| \vec v^{_H}\| \|\vec u\| \sin\psi$ with $\psi \in [0,\pi/2]$ for $\kappa = \kappa(A)$ implies:
\begin{equation}\label{eq:thrm2-40}
\frac{\|\vec v^{_H}A\| \|A^{-1} \vec u\|}{\|\vec v^{_H}\|\|\vec u\|} \leq \frac{\sin\psi}{\sin\phi} = \frac{1}{2}\left[{(\kappa+\kappa^{-1})+(\kappa-\kappa^{-1})\cos\psi}\right]. \nonumber
\end{equation}
Now, for any non-null vectors $\vec v$ and $\vec u$, and any positive definite matrix $M=A A^{-1}$:
\begin{equation}\label{eq:thrm2-42}
\frac{(\vec u^{_H}M \vec u)(\vec v^{_H}M^{-1} \vec v)}{(\vec u^{_H} \vec u)(\vec v^{_H} \vec v)} \leq \frac{\sin^2\psi}{\sin^2\phi} = \frac{\left[{(\kappa^2+1)+(\kappa^2-1)\cos\psi}\right]^2}{4\kappa^2} \nonumber 
\end{equation}
where  $\kappa^2 = \kappa(M)$.
Re-arranging terms then gives this equivalently as Theorem \ref{thm:bah}\footnote{For $\psi=\pi/2$, and $\vec u= \vec v$, this recovers the Kantorovich inequality derived in \ref{app:Thrm1}}:
\begin{equation}\label{eq:thrm2-42}
\frac{(\vec u^{_H} \vec v)^2}{(\vec u^{_H}M \vec u)(\vec v^*M^{-1} \vec v)}\geq  \frac{4}{\kappa_{\psi} +2 + \kappa_{\psi}^{-1} } ~\mbox{for}~ \kappa_{\psi} = \kappa(M) \tfrac{1+\sin \psi}{1 - \sin \psi}.\nonumber
\end{equation}

\section{Minimax degeneracy} \label{app:minima}

When the mean variance optimisation problem is unconstrained, then the direction of the optimised weights vector $\ve \theta^*$ in a mean variance optimisation procedure can be given by $\ve \theta^* \propto \Sigma^{-1} \ve \alpha$ and the quantity $\cos(\phi)$ is shown to be bounded from below when using that the covariance matrix is positive semi-definite and can be factored as $\Sigma = S' S$ in terms of symmetric matrices S and choosing that  $\vec v = S \ve \alpha$ in the Kantorovich inequality (Theorem \ref{thm:kant} in Section \ref{ssec:unconstrained}):
\begin{align}\label{eq:G1}
\cos^2{\phi} &=\frac{(\ve \alpha' \ve \theta^*)^2}{({\ve \alpha}' \ve \alpha)({\ve \theta^*}' \ve \theta^*)} \nonumber \\
&= 
\frac{(\vec v' \vec v)}
{(\vec v' \Sigma \vec v) (\vec v' \Sigma^{-1} \vec v)}
\geq \left[ {\frac{\sqrt{\rho_{n}\rho_{1}}}{\frac{1}{2}(\rho_{n}+\rho_{1})}}\right]^2. 
\end{align}
The lower bound quantity $\frac{\sqrt{\rho_{n}\rho_{1}}}{\frac{1}{2} (\rho_{n}+\rho_{1})}$ is the minimax degeneracy (or condition number) of the matrix $\Sigma$. The $\ve \alpha$-weight angle can be thus be estimated from the covariance matrix $\Sigma$ with eigenvalues $\rho_{n}$ and eigenvectors $\vec q_{n}$. This presents the worst case angle that can be achieved when attempting to align the direction of the optimal risky portfolio with that of the direction of expected returns. Equality can only be achieved in Equation (\ref{eq:G1}) when:
\begin{equation}\label{eq:G2}
\ve \alpha = \pm \sqrt{\rho_{1}}\vec q_{1} \pm \sqrt{\rho_{n}}\vec q_{n}~\mbox{and}~\ve \theta^* = \pm \frac{\vec q_{1}}{\sqrt{\rho_{1}}} \pm \frac{\vec q_{n}}{\sqrt{\rho_{n}}}.
\end{equation}
This is visualised in Figure \ref{fig:minmax}.
This worst case angle of alignment can be seen to be problematic as the portfolio is largely invested along the lowest volatility principal component $\vec q_n$. This implies a potentially significant loss of investment performance due to estimation error (See \ref{app:minima}). 

%% use tikz
\begin{figure}[h!] %\sidecaption
    \centering
    \usetikzlibrary{quotes,angles}
    \usetikzlibrary{arrows}
        \begin{tikzpicture} 
            scale=5,
            % axes 
            \draw[-stealth, thick] (0,0) -- (2.5,0) node[below] {$\vec q_{n}$ (min)};
            \draw[thick, -stealth] (0,0) -- (0,2.5) node[right] {$\vec q_{1}$ (max)};
            % angles
            \draw[very thick, -stealth] (0,0) coordinate (b) -- (0.7,2.3) coordinate (c) node[below right] {$\ve \alpha=\sqrt{\rho_{1}} \vec q_1 +\sqrt{\rho_n} \vec q_n$};
            \draw[very thick, -stealth] (0,0) -- (1,1.3) coordinate (a) node[ right] {$\ve \theta^*=\dfrac{\vec q_1} {\sqrt{\rho_1}}+\dfrac{\vec q_n}{\sqrt{\rho_n}}$}
            pic["$\phi$", draw=black, <->, angle eccentricity=1.2, angle radius=1.2cm]
            {angle=a--b--c};
            \draw[very thick, -stealth] (0,0) -- (1.5,0.5) coordinate (d) node[right] {$\ve \theta_z^*=\dfrac{\vec q_1} {\sqrt{\eta \rho_1}}+\dfrac{\vec q_n}{\sqrt{\rho_n}}$}
            pic["$\varphi$", draw=black, <->, angle eccentricity=1.25, angle radius=0.95cm]
            {angle=d--b--c};
        \end{tikzpicture}
    %\end{subfigure}
    \caption{The constrained $\ve \alpha$-weight angle and maximum and minimum  eigenvalues, $\rho_1$ and $\rho_n$, with associated eigenvectors $\vec q_1$ and $\vec q_n$, where $\ve \alpha' \ve \theta = |\ve \alpha| |\ve \theta| \cos(\varphi)$. Here $\ve \theta$ is the optimal portfolio control, and $\ve \alpha$ the expected asset return vector. This visualises the extent to which the optimal portfolio is typically invested aligned with the lowest volatility direction $\vec q_n$.}
    \label{fig:minmax}
\end{figure}

Similarly, when the mean variance optimisation problem is linearly constrained, in particular when we get solutions VI and VII (from Table \ref{tab:linearcons}), the direction of the optimal portfolio weight vector $\ve \theta^*$ is now given by $\ve \theta_z^* \propto \Sigma^{-1} \vec z$ in terms of a vector $\vec z$ that does not in general align with the direction of the returns $\ve \alpha$ because it now includes dependency on the constraints and we use $\vec y = S \vec z$.:
\begin{align}\label{eq:G3}
\cos^2 \varphi &= \frac{(\ve \alpha' \ve \theta_z^*)^2}{(\ve \alpha' \ve \alpha)({\ve \theta_z^*}' \ve \theta_z^*)}  \nonumber \\ 
&= \frac{(\vec v' \vec z)}
{(\vec v' \Sigma \vec v) (\vec z' \Sigma^{-1} \vec z)}
\geq \frac{4 \kappa_{\psi}}{(\kappa_{\psi} + 1)^2}
\end{align}
Here $\psi$ is $0\geq\psi\ge\frac{\pi}{2}$ and smaller than the angle $\varphi$
\begin{equation}
    \kappa_{\psi} =\frac{\rho_{1}}{\rho_{n}}\frac{(1+\sin(\psi))}{(1-\sin(\psi))}. 
\end{equation}
By setting $\eta = \frac{(1+\sin(\psi))}{(1-\sin(\psi))}\ge1$ we can calculate that the worst case angle for alignment with the expected returns from equality with the lower bound achieved when:
\begin{equation}\label{eq:G4}
\ve \alpha = \pm \sqrt{\eta \rho_{1}}\vec q_1 \pm \sqrt{\rho_{n}}\vec q_n~\mbox{and}~\ve \theta^*_z = \pm \frac{\vec q_1}{\sqrt{\eta \rho_1}} \pm \frac{\vec q_n}{\sqrt{\rho_n}}.
\end{equation}
Again this is problematic (See \ref{app:minima}). However, a key nuance is that optimal portfolio lower bound in gearing constrained optimisations, optimisations VI and VII (from Table \ref{tab:linearcons}), is given by equation (\ref{eq:G4}) from \citet{bauerhouseholder1960}, but that the key lower bound for the optimal risky portfolio remains that given in equation (\ref{eq:G2}) as governed by the Kantorovich inequality \citep{kant21964,kant11985}. This is the key insight that allows the refined shrinkage method to be defined: 1.) constrain the angle as in the unconstrained optimisation using the method of \cite{goltsandjones2009}, and then 2.) include the linear constraint restrictions. This is also why exclusively deciding between either the optimal mean-variance portfolio or the equally weighted benchmark is inconsistent and sub-optimal irrespective of the level of estimation uncertainty (similarly confirmed under normally distributed numerical simulation in Section \ref{sec:simports}). 

\section{Two-fund Separation} \label{app:2fund}

The two fund separation theorem derived by \citet{markowitz1952} states that: i.) the optimal portfolio exists, ii.) it is unique for any given return expectation level $\alpha_p$, and iii.) it can be separated into two distinct portfolios $\ve a$ and $\ve b$ from which all optimal portfolios $\ve \theta^*$ can be generated:
    $\ve \theta^* = (1-\alpha_p) \vec a + \alpha_p (\vec a + \vec b)$.
    
This is can proved by first considering two new optimal portfolios $\ve g$ and $\ve h$. Then let $\ve q$ be another optimal portfolio such that $\ve q = \mu \ve g + (1-\mu) \ve h$ for a real number $\mu$.

Now $\alpha_g \neq \alpha_h$, because they are distinct portfolios. So, there exists an unique solution to the equation: $\alpha_q = \mu \alpha_g + (1-\mu)\alpha_h$. Consider another portfolio $\ve p$ with weights $(\mu , (1-\mu))$ which has invested in portfolio's $\ve g$ and $\ve h$, it will satisfy: $\ve p = \mu(\ve a + \ve b \alpha_g) + (1-\mu)(\ve a + \ve b \alpha_h)$, so that $\ve p = \ve a + \ve b \alpha_q$. Therefore, $\ve p = \ve q$; and the optimal portfolio is unique and exists. Further details relating to fund separation theorems can be seen in \citet{ingersoll1987}.

All investors who choose portfolios by examining only mean and variance can be satisfied by holding different combinations of only a few (in this case 2) portfolios regardless of their preferences. All of the original assets, therefore, can be purchased by just two portfolios, and the investors can then just buy them in various ratios. 

\section{Maximum return portfolio} \label{app:maxretgear}

To find an investment portfolio with maximum return at a given level of portfolio variance with a gearing constraint we want to solve:
\begin{equation}\label{eq:mrg1}
    \arg \max_{\ve\theta} \left\{ {\ve \theta' \ve \alpha} \right\} ~\mathrm{s.t.}~\sigma_p^2 = \ve \theta' \Sigma \ve \theta \le \sigma_0 \mbox{ and }  \vec 1' \ve \theta = g_0.
\end{equation}
which can be reduced to the following Lagrangian with two Lagrange multipliers $\ve \lambda = (\lambda_1,\lambda_2)$ but with a quadratic constraint and a linear constraint:
\begin{equation}\label{eq:mrg2}
L =  - \ve \theta'\ve \alpha - \ve \lambda' \left( {\begin{matrix} \sigma_p^2 - \ve \theta' \Sigma \ve \theta \\ g_0 - \vec 1' \ve \theta \end{matrix} } \right).
\end{equation}
It should be expected here that picking a gearing $g_0$ necessarily fixes the variance of the portfolio. However there are multiple portfolios with the different portfolio variances, but the same leverage. We would like the minimum variance portfolio at a particular gearing $g_0$. This is then a multi-objective function optimisation. Hence we rather find the targeted variance first, and then set that variance to provide a portfolio with the required leverage. 

Here we want to ensure that there is an upper bound on portfolio risk: $\sigma^2_p \le \sigma^2_0$ and we know that a choice of $\sigma_p^2$ will set the portfolio gearing. Hence we rather solve the well known reduced problem with a single Lagrange multiplier $\lambda$ and then fix the leverage:
\begin{equation}\label{eq:mrg15}
L =  - \ve \theta'\ve \alpha - \lambda (\sigma_p^2 - \ve \theta' \Sigma \ve \theta).
\end{equation}
The first order (Kuhn-Tucker) conditions are: $L_{\ve \theta} = -\ve \alpha + \lambda 2 \Sigma \ve \theta = 0$, and $L_{\lambda} = \sigma_p^2 - \ve \theta' \Sigma \ve \theta = 0$. The second order conditions are $L_{\ve \theta \ve \theta} = 2 \lambda \Sigma \vec 1 \ge \vec 0 \implies  \lambda \ge 0$. 

Multiplying the first order condition $L_{\ve \theta}$ by $\ve \theta'$ and $\Sigma^{-1}$ to find:
\begin{eqnarray}
    0 &=& - \ve \theta' \ve \alpha + (2 \lambda) \ve \theta' \Sigma \ve \theta,\label{eq:mrg19}\\
    0 &=& - \Sigma^{-1} \ve \alpha + (2 \lambda) \ve \theta.\label{eq:mrg20}
\end{eqnarray}
We factor the optimal portfolio out of equation (\ref{eq:mrg20}):
\begin{equation}
    (2 \lambda) \ve \theta^{*} = {\Sigma^{-1} \ve \alpha}. \label{eq:mrg9}
\end{equation}
We now make use of the following substitutions 
\begin{align}
    A &= \vec 1' \Sigma^{-1}\vec 1 \label{app:eqn:defA}\\
    B &= \ve \alpha' \Sigma^{-1}\vec 1 = \vec 1'\Sigma^{-1}\ve \alpha \label{app:eqn:defB}\\
    C &= \ve \alpha'\Sigma^{-1}\ve \alpha \label{app:eqn:defC}\\
    D &= AC - B^2. \label{app:eqn:defD}
\end{align}
Here $D$ is positive, by the Cauchy-Schwarz inequality, since we have assumed that $\Sigma$ is non-singular and all assets do not have the same mean: $\ve \alpha \neq \frac{|\ve \alpha|}{n} \vec 1$. If all means were the same, then $D = 0$, and this problem has no solution.

Multiply equation (\ref{eq:mrg20}) from the left by $\ve \alpha'$ to find:
\begin{equation} \label{eq:mrg22}
  C = (2 \lambda) \alpha_p.   
\end{equation}
To find $\lambda$ we now substitute the portfolio expected returns and risk into equation (\ref{eq:mrg19}) to find $\alpha_p  = (2 \lambda)^2 \sigma^2_p$ and then substitute $\alpha_p$ into equation (\ref{eq:mrg22}) to find:
\begin{equation}
    (2 \lambda)^2 \sigma^2_p = C.
\end{equation}
Which gives the solution to optimisation I when substituted into equation (\ref{eq:mrg9}):
\begin{equation} \label{eq:mrg23}
    \ve \theta^*_1 = \frac{\sigma_p}{\sqrt{C}} \Sigma^{-1} \ve \alpha. 
\end{equation}
To set the gearing we multiply (\ref{eq:mrg23}) from the left by $\vec 1'$ to find that
\begin{equation}
    g_0 = \sigma_p \frac{B}{\sqrt{C}}. 
\end{equation}
The solution to optimisation V in terms of gearing is:
\begin{equation}
    \ve \theta^*_5 = g_0 \frac{1}{B} \Sigma^{-1} \ve \alpha = g_0 \ve \theta_{\alpha}.
\end{equation}
This is just a gearing optimal risky portfolio. For this reason optimisations I and V are treated as equivalent as they have the same investment directions. The ratio $B/A$ is just the ratio of the variance of optimal risky portfolio to the variance of global minimum variance portfolio.  

\section{Minimum risk portfolio} \label{app:gearing}

We want to find an investment portfolio with minimum variance and a targeted return with the additional gearing constraint, hence we want to solve for:
\begin{equation}\label{eq:14}
    \arg \min_{\ve\theta} \left\{ {\frac{1}{2}\ve \theta'\Sigma\ve \theta} \right\} \mathrm{ s.t. }~\ve \alpha' \ve \theta = \alpha_p \ge \alpha_0~\mbox{and}~\vec 1' \ve \theta = g_0.
\end{equation}
which can be reduced to the following Lagrangian with two Lagrange multipliers $\ve \lambda = (\lambda_1,\lambda_2)$:
\begin{equation}\label{eq:15}
L =  \frac{1}{2} \ve \theta'\Sigma\ve \theta - \ve \lambda' \left( {\begin{matrix} \ve \alpha' \ve \theta - \alpha_p \\ \vec 1' \ve \theta - g_0 \end{matrix} } \right).
\end{equation}
From the Kuhn-Tucker conditions
\begin{align}
    L_{\ve \theta} &= \Sigma \ve \theta - \ve \lambda' \left[ {\begin{matrix} \ve \alpha \\ \vec 1 \end{matrix}} \right] = 0, \label{eq:16} \\
    L_{\ve \lambda} &= \left[ {\begin{matrix} {\ve \alpha'\ve \theta - \alpha_p} \\ {\vec 1'\ve \theta - g_0} \end{matrix}} \right] = 0 \label{eq:16.5}
\end{align}
Solving for $\ve \theta$ in terms of $\ve \lambda$ from (\ref{eq:16})
\begin{equation}\label{eq:17}
    \ve \theta^*_6 = \Sigma^{-1} \ve \lambda \left[ {\begin{matrix} \ve \alpha \\ \vec 1 \end{matrix}} \right] = \lambda_1 \Sigma^{-1}\ve\alpha + \lambda_2 \Sigma^{-1}\vec1
\end{equation}
We now solve for $\ve \lambda$  by multiplying (\ref{eq:17}) first by $\vec 1'$ and then by $\ve \alpha'$ and then substituting these into equation (\ref{eq:16.5}):
\begin{align}
    g_0 = \lambda_1 \vec 1' \Sigma^{-1} \ve \alpha + \lambda_2 \vec 1' \Sigma^{-1} \vec 1 \label{eq:18} \\ 
    \alpha_p = \lambda_1 \ve \alpha' \Sigma^{-1} \ve \alpha +\lambda_2 \ve \alpha' \Sigma^{-1} \vec 1 \label{eq:18.5}
\end{align}
We again make use of the following substitutions from equations (\ref{app:eqn:defA}),(\ref{app:eqn:defB}),(\ref{app:eqn:defC}) and (\ref{app:eqn:defD}). Using these we simplify equations (\ref{eq:18}) and (\ref{eq:18.5}) to:
\begin{align}
    g_0 = \lambda_1 B + \lambda_2 A \label{eq:20}, \\
    \alpha_p = \lambda_1 C + \lambda_2 B. \label{eq:20.5}
\end{align}
From (\ref{eq:20.5}) above, we can see that
\begin{equation}\label{eq:21}
    \lambda_2 = \frac{\alpha_p - \lambda_1 C}{B}
\end{equation}
And by substituting (\ref{eq:21}) above into (\ref{eq:20})
\begin{equation}\label{eq:22}
   g_0 = \lambda_1 B + \frac{A}{B}({\alpha_p  - \lambda_1 C})
\end{equation}
Solving for $\lambda_1$:
\begin{equation}\label{eq:23}
   \lambda_1 = \frac{\alpha_p A - g_0B}{D}.
\end{equation}
Substituting (\ref{eq:23}) into (\ref{eq:21})
we then solve for $\lambda_2$:
\begin{equation}\label{eq:24}
   \lambda_2 = \frac{\alpha_p A - g_0 B}{D}
\end{equation}
Finally, from (\ref{eq:23}) and (\ref{eq:24}) as substituted into (\ref{eq:17}) we find
\begin{equation}\label{eq:25}
   \ve \theta^*_6 = \left( \frac{\alpha_pA-g_0B}{D} \right) \Sigma^{-1} \ve\alpha + \left( {\frac{g_0C -\alpha_pB}{D}} \right) \Sigma^{-1} \vec1 
\end{equation}
We can then put this into the form $\ve \theta \propto \Sigma^{-1} \vec z$, where $\vec z$ is the same dimension as $\ve \alpha$:
\begin{equation}\label{eq:25.5}
   \ve \theta^*_6 = \Sigma^{-1}\left ( \left( \frac{\alpha_pA-g_0B}{D} \right) \ve\alpha + \left( {\frac{g_0C -\alpha_pB}{D}} \right) \vec1 \right ). \nonumber
\end{equation}
This can be reduced further to find the global minimum risk portfolio associated with a particular value of the gearing parameter $g_0$, this allows us to define a second efficient frontier, as a function of $g_0$, by eliminating $\alpha_p$ by enforcing that $\partial_{\alpha_p} \sigma^2_p = 0$. We are effectively putting a lower bound on the portfolio returns $\alpha_0$ by associating this with the unleveraged global minimum variance portfolio. 

\section{Mean-variance portfolio} \label{app:mvgearing}

The gearing constrained mean-variance optimal portfolio choice problem is given by the optimisation:
\begin{align}\label{app:eq:mvgearing1}
    \argmax_{\ve \theta} \left\{ { \ve \theta'\ve  \alpha - \cfrac{\gamma}{2} \ve \theta'\Sigma \ve \theta }\right\} \quad \mbox{s.t} \quad \vec 1' \ve \theta = g_0.
\end{align}
It has a Lagrangian with a single Lagrage multiplier:
\begin{equation} \label{app:eq:mvgearing2}
L = - \ve \theta' \ve \alpha + \frac{\gamma}{2} \ve \theta' \Sigma \ve \theta - \lambda (\ve \theta' \vec 1 - g_0). 
\end{equation}
Solving the first order optimality conditions (the Kuhn-Tucker equations) we find:
\begin{align}
    L_{\ve \theta} &= -\ve \alpha + \gamma \Sigma \ve \theta - \lambda \vec 1 = \vec 0, \label{app:eq:mvgearing3} \\
    L_{\ve \lambda} &= \ve \theta' \vec 1 - g_0 = 0 \label{app:eq:mvgearing4}
\end{align}
Using equation (\ref{app:eq:mvgearing3}) to solve for the optimal control $\ve \theta^*$ and then substituting this into equation (\ref{app:eq:mvgearing4}) to eliminate the Lagrange multiplier, to then find the optimal portfolio:
\begin{align}\label{eq:mvgearing5}
    \ve \theta^{*}_7 = \left({g_0 - \frac{1}{\gamma} \vec 1'\Sigma^{-1}\ve  \alpha }\right) \frac{\Sigma^{-1}\vec 1}{\vec 1'\Sigma^{-1}\vec 1} + \frac{1}{\gamma}\Sigma^{-1}\ve  \alpha. \nonumber
\end{align}

\section{Optimal risky portfolio} \label{app:risky}

To find an investment portfolio that maximises the Sharpe ratio $\alpha_p/\sigma_p$ but with a gearing constraint and setting the risk free rate to zero; we want to solve:
\begin{equation}\label{eq:risky1}
    \arg \max_{\ve\theta} \left\{ {\frac{\ve \theta' \ve \alpha}{\sqrt{\ve \theta' \Sigma \ve \theta}}} \right\} ~\mathrm{s.t.}~ \vec 1' \ve \theta = g_0.
\end{equation}
This can be reduced to a Lagrangian with a single Lagrange multiplier:
\begin{equation}\label{eq:risky2}
L =  - \frac{\ve \theta'\ve \alpha}{\sqrt{\ve \theta' \Sigma \ve \theta}} - \lambda ( \ve \theta' \vec 1 - g_0).
\end{equation}

From the first order conditions
\begin{align}
    L_{\ve \theta} &=  - \frac{\ve \alpha}{(\ve \theta' \Sigma \ve \theta)^{\frac{1}{2}}} + \frac{\ve \theta' \ve \alpha}{(\ve \theta' \Sigma \ve \theta)^\frac{3}{2}} \Sigma \ve \theta - \lambda \vec 1 = \vec 0. \label{eq:risky3} \\
    L_{\ve \lambda} &= \ve \theta' \vec 1 - g_0 = 0. \label{eq:risky4}
\end{align}
We can multiply equation (\ref{eq:risky3}) by $\ve \theta'$ and using equation (\ref{eq:risky4}) we find that the Lagrange multiplier: $\lambda g_0 = 0$ and hence that $\lambda=0$ for $g_0\neq0$.

Using this and multiplying equation (\ref{eq:risky3}) from the left by $\Sigma^{-1}$ we can show that:
\begin{equation} \label{eq:risky5}
 \ve \theta^*_8 = \left({\frac{\ve \theta' \ve \alpha}{\sqrt{\ve \theta' \Sigma \ve \theta}}}\right)^{-1} \Sigma^{-1} \ve \alpha.
\end{equation}
This is convenient because we can now multiply this from the left by the transposed unit vector $\vec 1'$ and use equation (\ref{eq:risky4}), to find that:
\begin{equation} \label{eq:risky6}
    g_{0} =  \left({\frac{\ve \theta' \ve \alpha}{\sqrt{\ve \theta' \Sigma \ve \theta}}}\right)^{-1} {\vec 1' \Sigma^{-1} \ve \alpha}.
\end{equation}
Substituting this into equation (\ref{eq:risky5}) this gives the optimal leveraged risk adjusted portfolio:
\begin{equation} \label{eq:risky7}
    \ve \theta^*_8 = g_0 \frac{\Sigma^{-1} \ve \alpha}{\vec 1' \Sigma^{-1} \ve \alpha} = g_0 \ve \theta_{\alpha}. 
\end{equation}
We note that optimisation I, II, III and IV are all special cases of general leverage Sharpe ratio optimisation in problem VIII -- they are all geared optimal risky portfolios.

\section{Portfolio diversity constraint} \label{app:quadcons}

The problem of finding a portfolio constrained to a given number of effective bets $n_0$ \citep{GALLUCCIO1998449,GABOR1999222} requires a quadratic constraint, and this constraint can been seen to act as a regularisation condition. Such an investment portfolio with minimum variance, gearing constrained, but with a quadratic constraints is the following problem:
\begin{equation}\label{eq:divcons}
    \arg \max_{\ve\theta} \left\{ {\ve \theta' \ve \alpha - \frac{\gamma}{2}\ve \theta'\Sigma\ve \theta} \right\} \mathrm{ s.t. }~~  \vec 1' \ve \theta = g_0 \mbox{ and }  \ve \theta' \ve \theta = \sfrac{1}{n_0}. \nonumber
\end{equation}
This is a special case of the problem of Quadratic Optimisation with Quadratic Constraint (QOQC) \citep{tuyhoaiphuong2007}. Here the Lagrangian has two Lagrange multipliers $\ve \lambda = (\lambda_1,\lambda_2)$:
\begin{equation}\label{eq:La3}
L =  - \ve \theta' \ve \alpha + \frac{\gamma}{2} \ve \theta'\Sigma\ve \theta - \ve \lambda' \left[ {\begin{matrix} \ve \theta' \ve \theta - \sfrac{1}{n_0} \\ \vec 1' \ve \theta - g_0 \end{matrix} } \right].
\end{equation}
The first order conditions are:
\begin{align}
    L_{\ve \theta} &= - \ve \alpha + {\gamma} \Sigma \ve \theta - \ve \lambda' \left[ {\begin{matrix} { \ve \theta} \\ {\vec 1 } \end{matrix}} \right] = 0, \label{eq:C3.1} \\
      L_{\ve \lambda} & = \left[ {\begin{matrix} {\ve \theta'\ve \theta - \sfrac{1}{n_0}}\\ {\vec 1'\ve \theta - g_0} \end{matrix}} \right] = 0. \label{eq:C3.3}
\end{align}
By multiplying equation (\ref{eq:C3.1}) by $\ve \theta$ and using the constraints (\ref{eq:C3.3}) we find that: $\lambda_2 = \alpha_p - \frac{1}{\gamma} \sigma^2_p - \frac{\lambda_1}{n_0}$.

The second order constraint gives a convexity condition:
\begin{equation}
    L_{\ve \theta \ve \theta} = {\gamma} \Sigma \vec 1 - \lambda_1 \vec 1  \ge \vec 0 \implies  \lambda_1 \ge \frac{\gamma}{n} \vec 1' \Sigma \vec 1. \label{eq:C3.2}
\end{equation}

Solving for $\ve \theta$ in terms of $\ve \lambda$ from equations (\ref{eq:C3.1}) and (\ref{eq:C3.3}) where for the identity matrix $I_n$ we find:
\begin{equation}\label{eq:C6}
    \ve \theta^* = \left( {\lambda_1 I_n + \gamma \Sigma} \right)^{-1} (\ve \alpha + \lambda_2 \vec 1).
\end{equation}
This is no longer of the form $\ve \theta^* \propto \Sigma^{-1} \vec x$ for some vector $\vec x$ but we can make the substitution:
\begin{equation} 
\tilde \Sigma = \left({\lambda_1 I_n + \gamma \Sigma} \right)
\end{equation}
to get the solution into a regularised form $\ve \theta^* \propto \tilde \Sigma^{-1} \vec x$. The regularised covariance matrix is still positive semi-definite and hence it can still be factorised into two symmetric matrices; the \cite{bauerhouseholder1960} extension of the Kantorovich inequality then remains valid with regards to the condition number of new covariance matrix $\tilde \Sigma$. So the $\ve \alpha$-angle is still bound below by the condition number. Hence improving the conditioning by aligning the resulting optimal portfolio with the expected returns makes sense.

\end{document}